\documentclass[iop,apj,numberedappendix,twocolappendix,appendixfloats]{emulateapj}

\usepackage{epsfig}
\usepackage{amssymb}
\usepackage{amsmath}
\usepackage{natbib}

\usepackage{setspace}

\def \kms{~\rm{km}~\rm{s}^{-1}}

\def \K{~\rm{K}}

\shorttitle{Recovery from Giant Eruptions in VMSs}
\shortauthors{A. Kashi, K. Davidson and R. M. Humphreys}

\begin{document}


\title{Recovery from Giant Eruptions in Very Massive Stars}

\author{
Amit Kashi\altaffilmark{1},
Kris Davidson\altaffilmark{1} and
Roberta M. Humphreys\altaffilmark{1}
}

\affil{
\altaffilmark{1}{Minnesota Institute for Astrophysics, University of Minnesota, 116 Church St. SE. Minneapolis, MN 55455, USA  ~~  \href{mailto:kashi@astro.umn.edu}{kashi@astro.umn.edu}}
}

\begin{abstract}

We use a hydro-and-radiative-transfer code to explore the behavior of a
very massive star (VMS) {\it after\/} a giant eruption -- i.e., following
a supernova impostor event. Beginning with reasonable models for
evolved VMSs with masses of $80~M_\odot$ and $120~M_\odot$, we simulate the change of state caused by a giant
eruption via two methods that explicitly conserve total energy:
1. Synthetically removing outer layers of mass of a few $M_\odot$ while reducing the
energy of the inner layers. 2. Synthetically transferring energy
from the core to the outer layers, an operation that automatically causes
mass ejection. Our focus is on the aftermath, not the poorly-understood
eruption itself. Then, using a radiation-hydrodynamic code in 1D with
realistic opacities and convection, the interior disequilibrium
state is followed for about 200 years. Typically the star develops a $\sim 400 ~\rm{km}~\rm{s}^{-1}$
wind with a mass loss rate that begins around
$0.1 ~M_\odot~\rm{yr^{-1}}$ and gradually decreases. This outflow is
driven by $\kappa$-mechanism radial pulsations.
The 1D models have regular pulsations but 3D models will probably be more chaotic.
In some cases a plateau in the mass-loss rate may persist about 200 years,
while other cases are more like $\eta$ Car which lost $>10~M_\odot$ and then had an abnormal mass loss rate for
more than a century after its eruption.
In our model, the post-eruption outflow carried more mass than the initial eruption.
These simulations constitute a useful preliminary reconnaissance
for 3D models which will be far more difficult.

\end{abstract}

\keywords{stars: evolution --- stars: mass loss --- stars: winds, outflows --- stars: variables: general --- hydrodynamics --- methods: numerical}

\linespread{1}

\section{Introduction}
\label{sec:intro}

Very massive stars (VMSs) with $M \gtrsim 80~M_\odot$ spend substantial fractions of their short lifetimes close to the Eddington limit.
After leaving the main sequence, they stay in the blue part of the HR diagram, left of the HD limit \citep{HumphreysDavidson1979}.
It is known from the excess of nitrogen in their ejecta that these stars have gone far into core hydrogen-burning via the CNO bi-cycle.
As the star lost most of its hydrogen in winds, and helium was dredged-up from the core, the outer layers are helium rich.
This is confirmed by observations of $\eta$ Car (\citealt{Davidson1986}; \citealt{Dufour1997}).
Being so close to the Eddington limit, VMSs are susceptible to instabilities and disturbances, both internally and from companion stars, which
can undermine the delicate balance between radiation pressure and gravity.
Perhaps the most exotic effect that can happen to such a fragile structure is a Giant Eruption, in which the VMS loses $1$--$20\%$ of its mass (e.g., \citealt{HumphreysDavidson1994}; \citealt{SmithOwocki2006}; \citealt{VanDykMatheson2012}).

These eruptions are sometimes called ``Supernova Impostors'',
and in some cases it is unclear whether the VMS underwent a giant eruption, or actually exploded as a supernova.
Some terminal explosions of VMSs are associated with type IIn SNe (e.g., \citealt{Gal-Yam2007}; \citealt{Kochanek2012}), but other types have also been suggested, such
as the so-called type IIb, an event which transitions from type II to Ib (\citealt{Groh2013}).

In that context it is interesting to mention SN~2011ht, a peculiar Type IIn supernova
with significant "impostor" characteristics in its spectrum (\citealt{Humphreys2012}; \citealt{Roming2012}).
An example with a more  complex history is SN~2009ip. In 2009 it experienced a
giant eruption initially labeled a SN event; but then a series of later
outbursts arose in 2011--2012, leaving researchers uncertain about
whether the terminal event had finally occurred or not
(\citealt{Drake2012}; \citealt{Fraser2013}; \citealt{Kashi2013}; \citealt{Mauerhan2013}, \citeyear{Mauerhan2014}; \citealt{Margutti2014}; \citealt{Pastorello2013}; \citealt{Prieto2013}; \citealt{SokerKashi2013}; \citealt{TsebrenkoSoker2013}; \citealt{Graham2014}).

Since this topic has not received much theoretical attention, far less is
known about the impostors and giant eruptions than about true supernovae.
The energy budget may allow either a core event or an instability in the
outer layers, or conceivably even both. Only a few examples have been
observed for sufficient lengths of time, the most famous case being
$\eta$ Car's Great Eruption in the nineteenth century. Some conjectural
models suggest interior mechanisms
(e.g., \citealt{StothersChin1997}; \citealt{Guzik1997}; \citealt{DwarkadasBalick1998}; \citealt{Langer1999}; \citealt{Baraffe2001}; \citealt{Shaviv2001}; \citealt{Gonzalez2004},\citeyear{Gonzalez2010}; \citealt{Woosley2007}; \citealt{ChatzopoulosWheeler2012}; \citealt{ShachamShaviv2012}; \citealt{Smithetal2011}; \citealt{Smith2013}),
while others employ tidal intervention by companion stars (\citealt{Cassinelli1999}; \citealt{KashiSoker2010}; \citealt{Kashi2010}).
The phenomenon {\it may} be physically related to LBV eruptions which
are much smaller. Using a nonlinear pulsation gas dynamics code,
\cite{GuzikLovekin2014} found that the Eddington limit can be crossed repeatedly in the
pulsation driving region of the stellar envelope, very likely generating
mass loss. The evolution of a massive star depends on its metallicity,
rotation, and mass-loss history as well as initial mass (e.g, \citealt{Heger2005}); and a nearby
companion can alter the composition and angular momentum distribution
-- maybe triggering an outburst.

A VMS can lose 1\% to 20\% of its mass in one or more giant eruptions (e.g., \citealt{HumphreysDavidson1994}; \citealt{Langer2012}; \citealt{DavidsonHumphreys2012a})
leaving it in a state of thermal and rotational disequilibrium.
\cite{Davidson2012} commented on possible time scales for the subsequent recovery.
For example, a change of state in the mass-loss rate is currently observed in $\eta$ Car (\citealt{Davidson2005}; \citealt{Mehner2012}, \citeyear{Mehner2014}),
suggesting that the recovery time is in the order of centuries, and the trend has been remarkably \emph{unsteady} \citep{Humphreys2008}.

The relevant timescale for the recovery is unclear. A thermal timescale depends on which layers are included, and we note that $\eta$ Car's
current luminosity amounts to $10^{50}$ ergs (half the amount released in the Great Eruption) in about 200 years.
However, the transport rate may have been faster soon after the event because thermal disequilibrium was worse then. Angular momentum, chemical
composition, and perhaps even entropy can be transported by turbulent viscosity, whose timescale in $\eta$ Car may be as short as 25 years \citep{Davidson2012}.
The important point is that recovery timescales may be anywhere in the range 10 to 300 years, based only on simplified reasoning without detailed simulations.
As $\eta$ Car, SN2009ip and other objects have shown, it may well be that during this recovery another giant eruption takes place.

In a previous experiment of giant eruption and recovery of an evolved star, \cite{HarpazSoker2009} took an evolved star model (after hydrogen in the core was depleted)
and modeled its response to a very high mass loss.
Their model had a convective core, then a radiative zone, followed by another convective shell and another outer radiative zone with very little mass.
It developed a steep entropy rise above the convective core.
To simulate an eruption, they subtracted mass from the evolved star at a constant rate of $1 ~M_\odot~\rm{yr^{-1}}$ for a duration of 20 years,
after which they reduced the mass-loss rate and traced the star for additional 200 years.
Their code, as they note, cannot conserve energy when removing the mass from the star.
In a response to the mass loss the star contracted (mainly due to the removal of the outer high entropy layers) and released a huge amount of gravitational energy that caused its luminosity to increase to values up to almost $10^9 L_\odot$.
Harpaz and Soker suggest that this will lead to a runaway mass loss.

In this paper we report on several experiments to model the recovery of a VMS from a Giant eruption.
The star is modeled from the core to the surface including the wind using the radiation transfer-hydro code \texttt{FLASH}.
This involves 12 or more orders of magnitude in density, and requires simulations with a high
level of refinement.
As a first step, to test our physical model and input parameters we describe the results of one-dimentional simulations which include the entire star and its wind.
They provide an initial view and guidance for 3D simulations.

$\eta$ Car will appear often in the following text, merely because it is the only post-eruption VMS that has been observed well.
Its luminosity, ejected mass, post-eruption wind and photometric record, etc., are known far better than any other supernova impostor
(\citealt{HumphreysMartin2012}; \citealt{Davidson2012}, and many references therein).
$\eta$ Car's Great Eruption in 1830--1860 may have been triggered by the companion star near periastron (\citealt{KashiSoker2010} and references therein),
but that possibility has little effect on the subsequent internal physics.
This paper concerns post-eruption VMS's in general, not specifically $\eta$ Car.

In principle we cannot be certain of $\eta$ Car's evolutionary state, because the surface wind does not directly represent the core structure.
The observed surface helium abundance shows that it is somewhat evolved.
The timescale for late stages is very short, and one might expect hydrogen should then be scarce.
Therefore the simplest as well as more probable choice is that it has not yet reached the helium burning stage --
but there is no known way to prove this.
These questions obviously require further investigation.
The mass loss has been unsteady, and for more than a century it was too strong to be explained by line driving (\citealt{Davidson2012}).

In section \ref{sec:methods} we describe our numerical model -- the derivation of the VMS model,
the simulation of the eruption, and the hydrodynamical run of its recovery.
The results are described in section \ref{sec:results}, for a number of studied cases.
Our summary and discussion are in section \ref{sec:Summary}.

\section{Methods}
\label{sec:methods}

\subsection{Obtaining a Pre-Eruption Model}
\label{sec:methods:LBV}

\begin{figure*}
\includegraphics[width=0.51\textwidth]{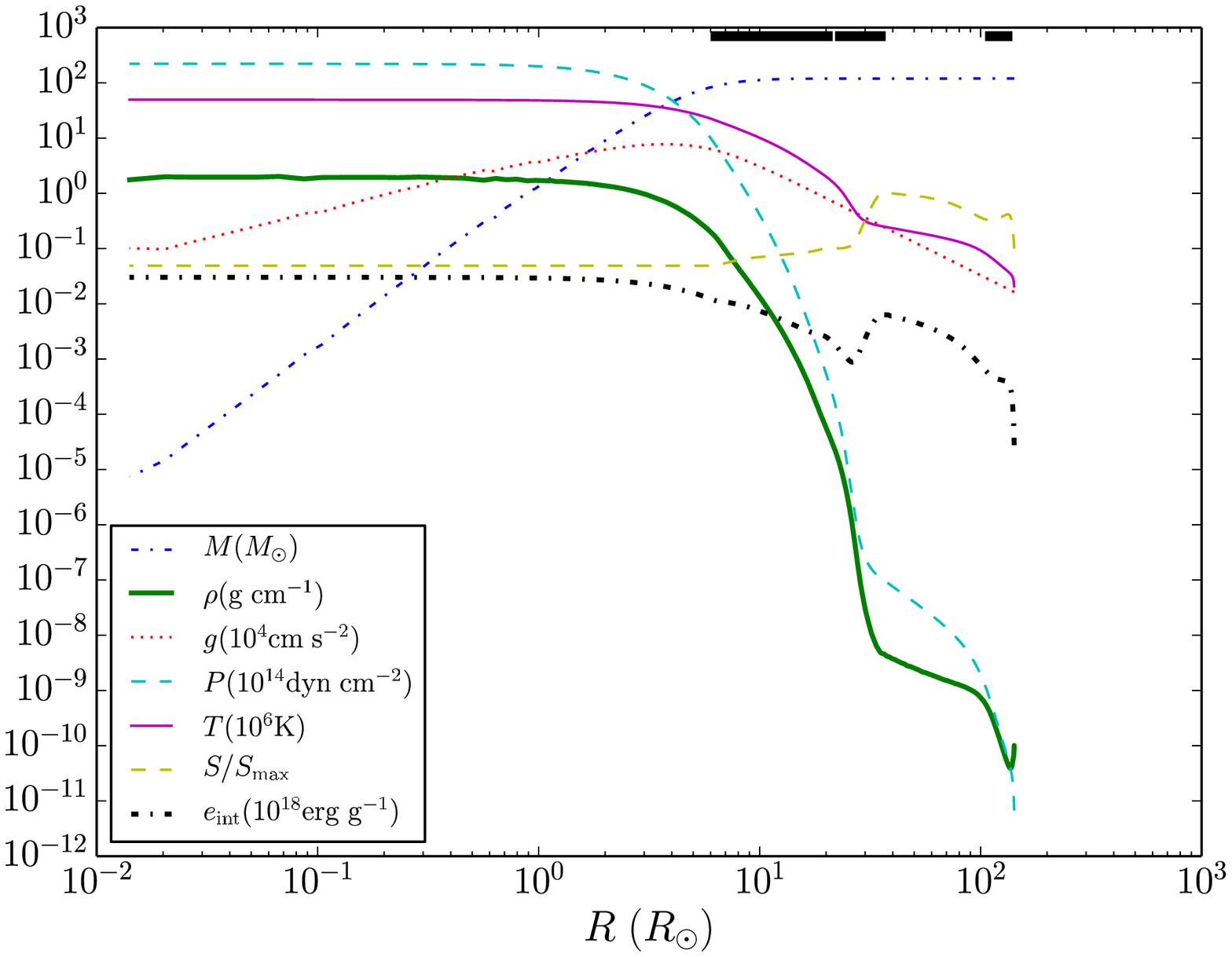}
\includegraphics[width=0.51\textwidth]{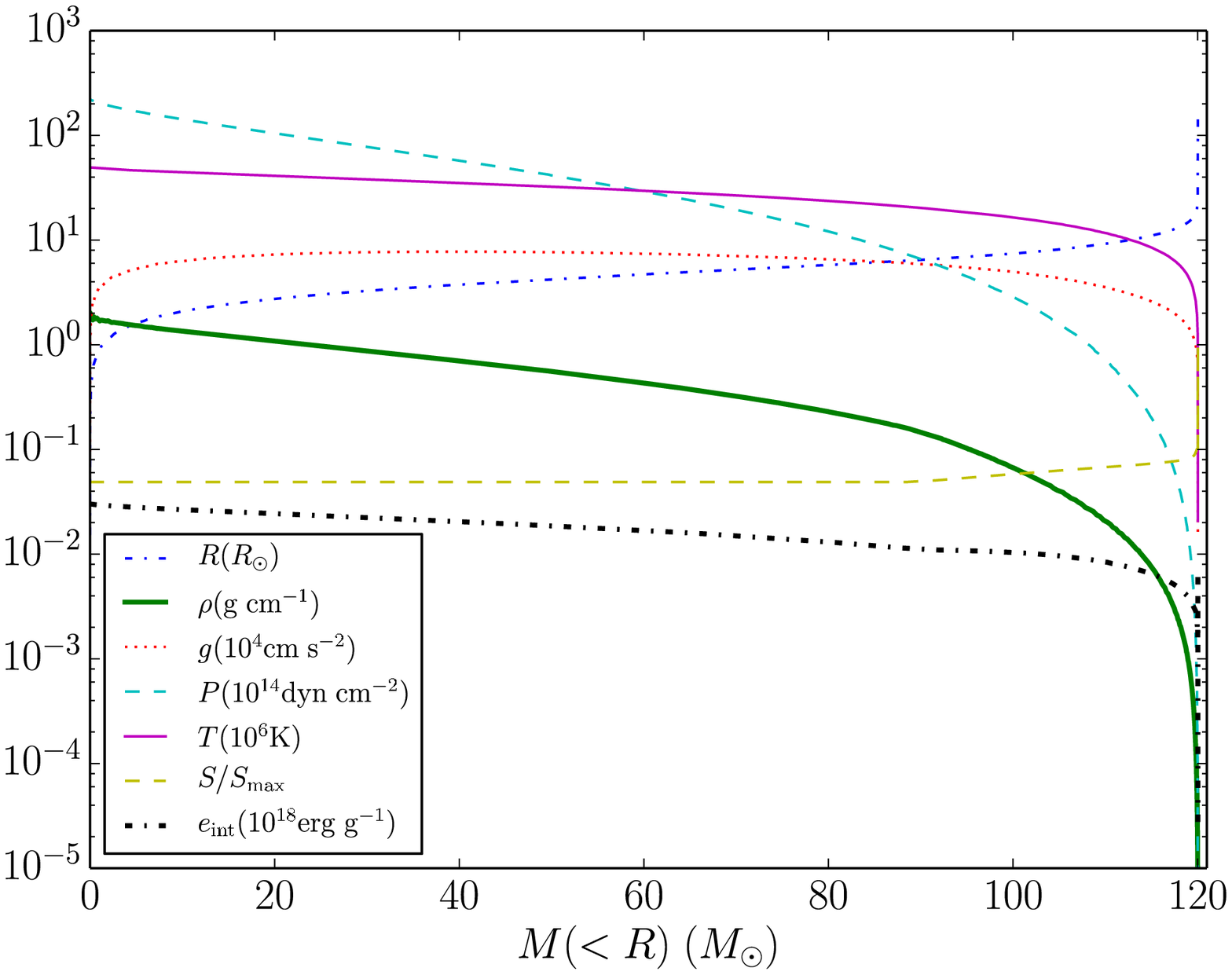}
\includegraphics[width=0.51\textwidth]{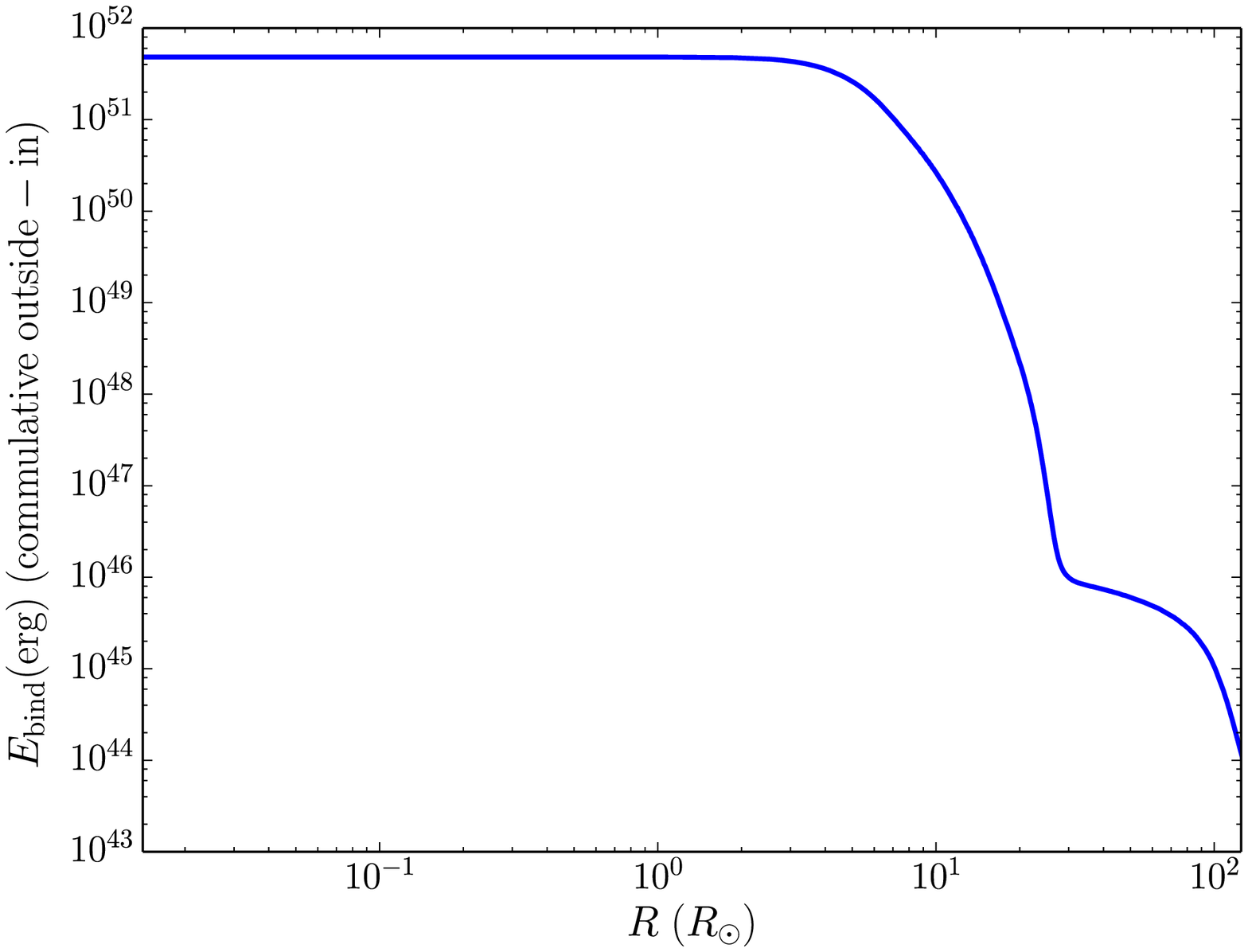}
\includegraphics[width=0.51\textwidth]{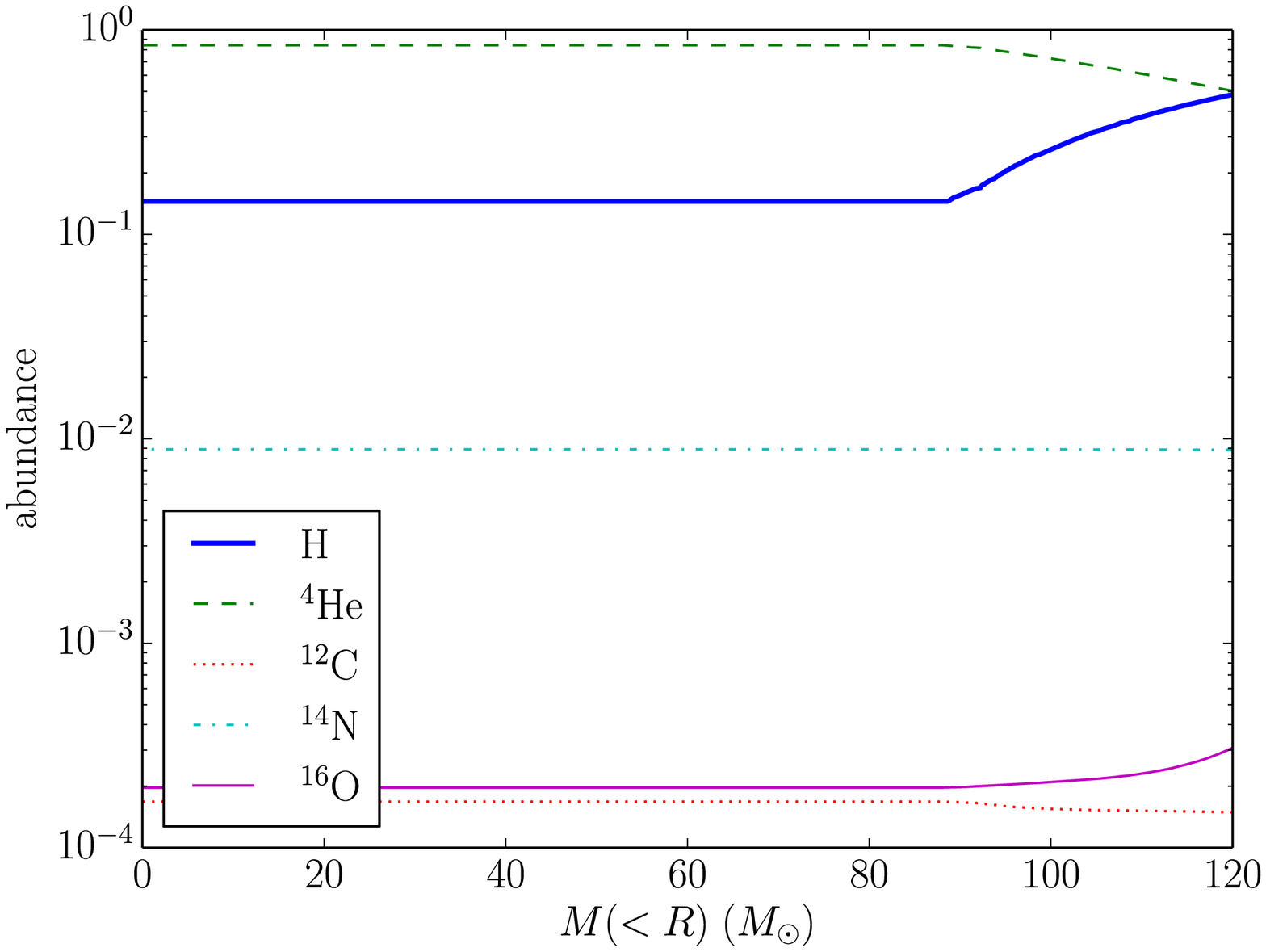}
\caption{
The properties of the $120~M_\odot$ very massive star (VMS) model, which represents an evolved star at a post main sequence evolutionary stage.
Upper left: The mass, density, gravitational acceleration, pressure, temperature, normalized entropy and specific internal energy as a function of stellar radius.
Upper right: The stellar properties as a function of mass inside $R$.
Lower left: The cumulative binding energy, integrated from the outer layer of the star inwards.
Lower right: The stellar composition as a function of stellar radius.
}
\label{fig:LBV1}
\end{figure*}

Our moderately evolved non-rotating VMS has
$M \approx 120 ~M_\odot$, $L \approx 4 \times 10^6 ~L_\odot$, and
its radius would correspond to $T_\mathrm{eff} \approx 20\,000 \K$ if
there were no opaque wind. In order to obtain a consistent internal
structure, we begin with a $170 ~M_\odot$ ZAMS star and evolve it
with the 1D \texttt{MESA} code (\citealt{Paxton2011},\citeyear{Paxton2013},\citeyear{Paxton2015}).
In order to obtain a post main sequence helium abundance $Y \sim 0.5$, suggested by the $\eta$ Car example,
much of the hydrogen-rich envelope must be ejected.
Since the conventional mass-loss prescriptions usually employed with \texttt{MESA} (e.g., De Jager, Vink, etc.)
cannot achieve this, we adopt the following simple behavior.
(1) Initially the star evolves away from the ZAMS with a mass loss
rate of $\dot{M}=10^{-5} ~M_\odot~\rm{yr^{-1}}$.
(2) When $T_\mathrm{eff}$ falls below $20\,000 \K$ (the bi-stability jump), the rate increases to
$\dot{M}=10^{-3} ~M_\odot~\rm{yr^{-1}}$.
Obviously this is somewhat arbitrary, but the same is true for all published VMS evolution models, since
their mass-loss behavior at that stage is very poorly understood.
The purpose of our \texttt{MESA} usage is merely to obtain a ``reasonable''
internal structure just before the VMS eruption.
Since a very high mass-loss rate tends to increase $T_\mathrm{eff}$, the above
procedure creates a negative feedback that ``locks'' the star at
$T_\mathrm{eff} \approx 20\,000 \K$.
At the time when its mass has fallen to $120 ~M_\odot$, the luminosity
is $L \approx 3.4 \times 10^6 ~ L_\odot$ and the radius is about
$153 ~R_\odot$. The helium fraction at the surface is then
$Y \approx 0.5$.
Helium burning has not begun yet, and it is burning hydrogen via the CNO bi-cycle.
This model is adequate for our numerical experiment, and is credible in the sense that it resembles the best-observed
eruptive VMS, $\eta$ Car.
Its structural properties are shown in Fig. \ref{fig:LBV1}.

\subsection{Simulating the results of an Eruption}
\label{sec:methods:Eruption}

The next stage is to simulate a giant eruption in the VMS. The basic
instability is not yet understood (indeed there may be more than one
mechanism), and our goal is not to model it but rather to explore the
aftermath. The only unquestionable facts are that a substantial mass
is ejected along with $10^{49}$ to $10^{51}$ ergs of energy, and that
total energy must be conserved. Fortunately the energy requirement is
only a small fraction of the star's total thermal energy. We use
two operationally different approaches.
(1) {\it Synthetically removing mass\/} in a timescale much shorter
than the relevant thermal timescale. In other words, we suddenly
truncate the star's mass distribution. Energy to unbind and lift the
ejecta is artificially extracted from mass layers that are not removed.
If the removal timescale is longer than a few weeks, most of the star
remains fairly close to hydrostatic equilibrium.
The main unknown is the radial distribution of the special energy-supply
fund, which depends on the nature of the instability.
In any case the outermost surviving layers have far less energy than
the deeper layers; so, after a short-lived redistribution process,
the behavior depends only moderately on the assumed distribution.
In the models described here, we simply assume that the thermal energy
decreases by a constant fraction per unit mass.

(2) {\it Extracting thermal energy from the core.\/} In this approach we
artificially move some thermal energy from the inner region to the outer
layers, without explicitly removing any mass.
The location where we deposit the energy is the radius which has
the same gravitational binding energy above,
so the energy is sufficient to free the layers above.
The resulting structure is far out of thermal equilibrium, and automatically ejects some mass.
Thermal energy is extracted from mass layers as done for method 1.
Hydrostatic equilibrium is not preserved at the beginning.

\begin{figure*}
\includegraphics[trim= 1.0cm 0.4cm 0.6cm 0.8cm,clip=true,width=0.5\textwidth]{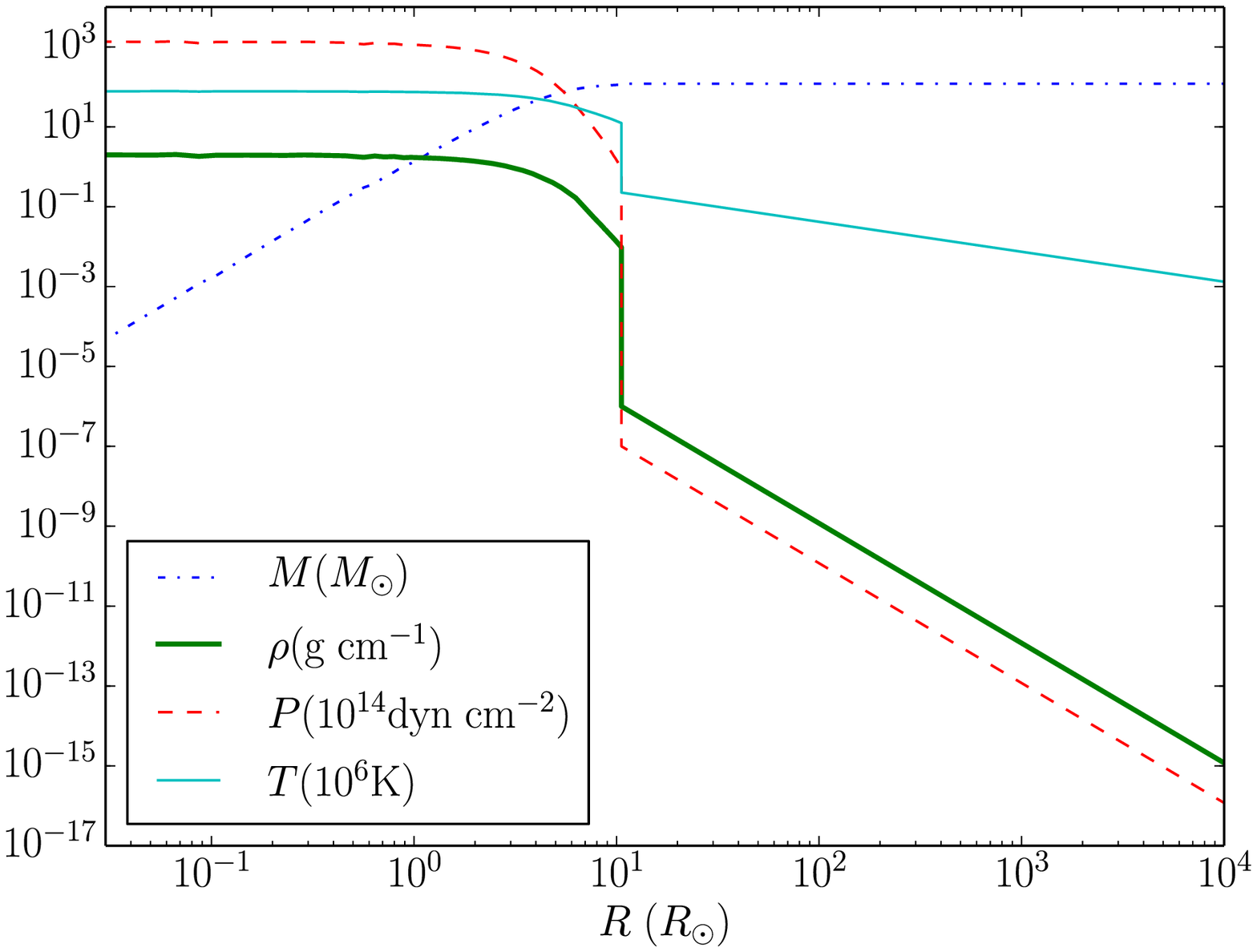}     
\includegraphics[trim= 1.0cm 0.4cm 0.6cm 0.8cm,clip=true,width=0.5\textwidth]{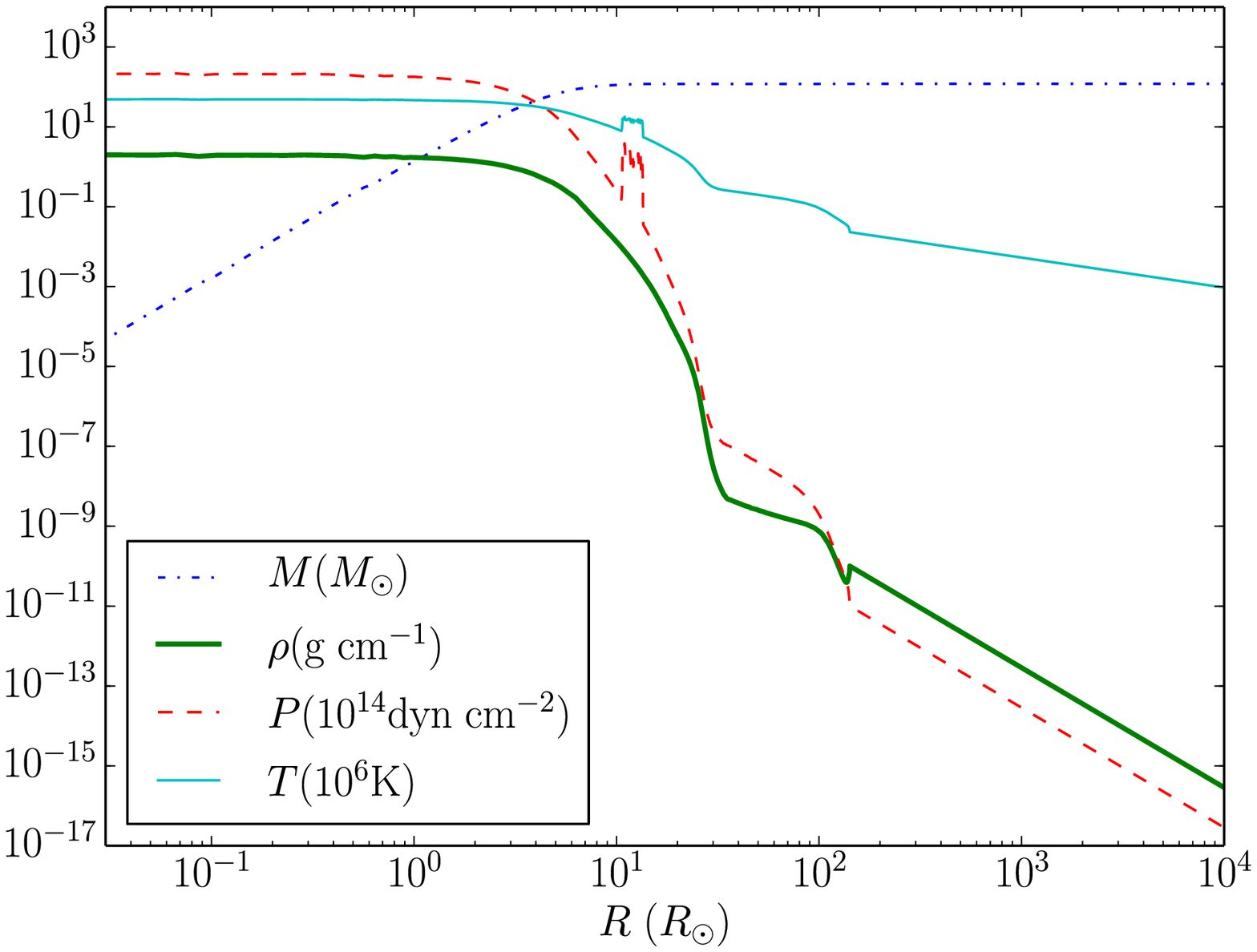}
\includegraphics[trim= 1.0cm 0.4cm 0.6cm 0.8cm,clip=true,width=0.5\textwidth]{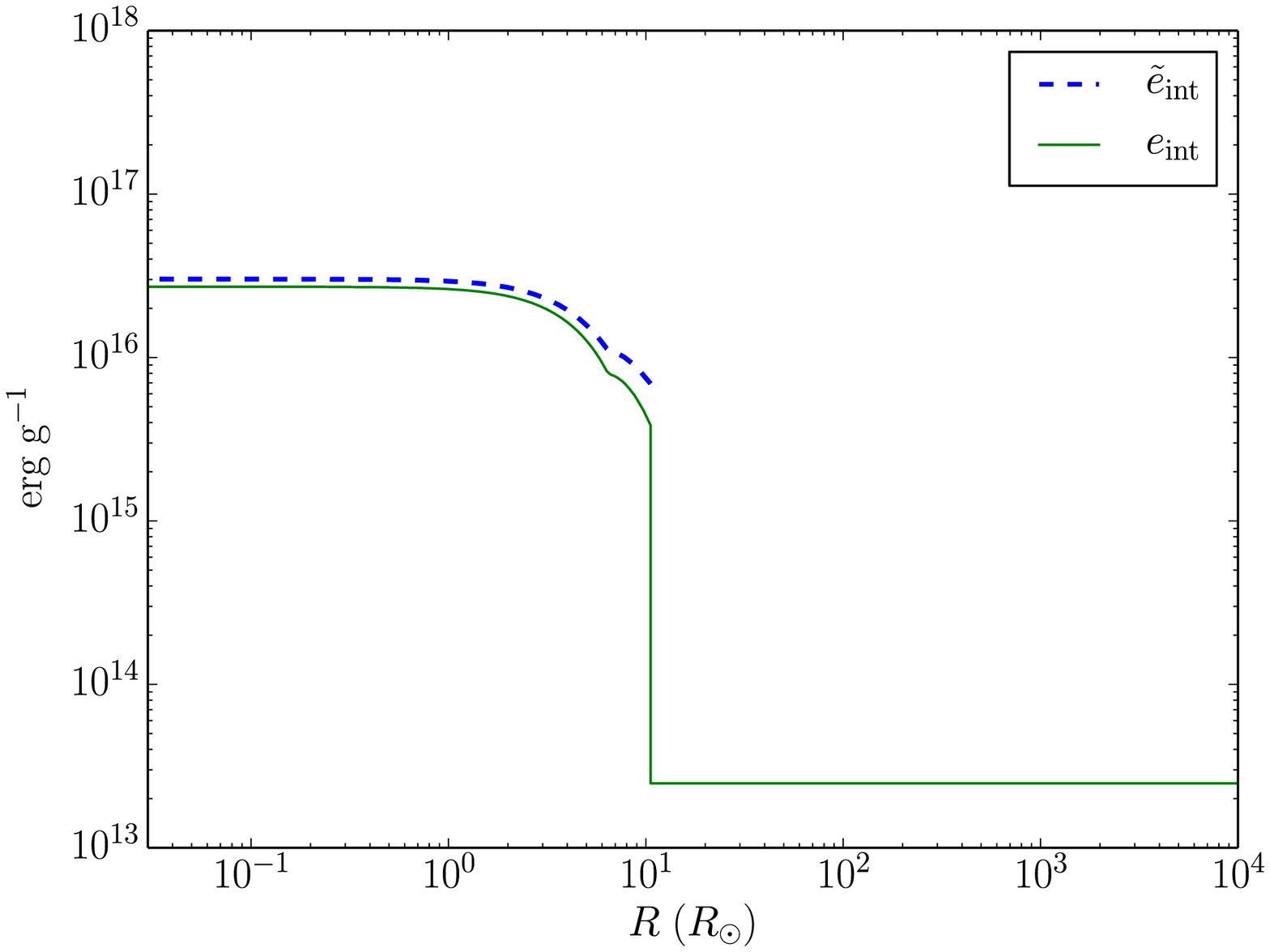}
\includegraphics[trim= 1.0cm 0.4cm 0.6cm 0.8cm,clip=true,width=0.5\textwidth]{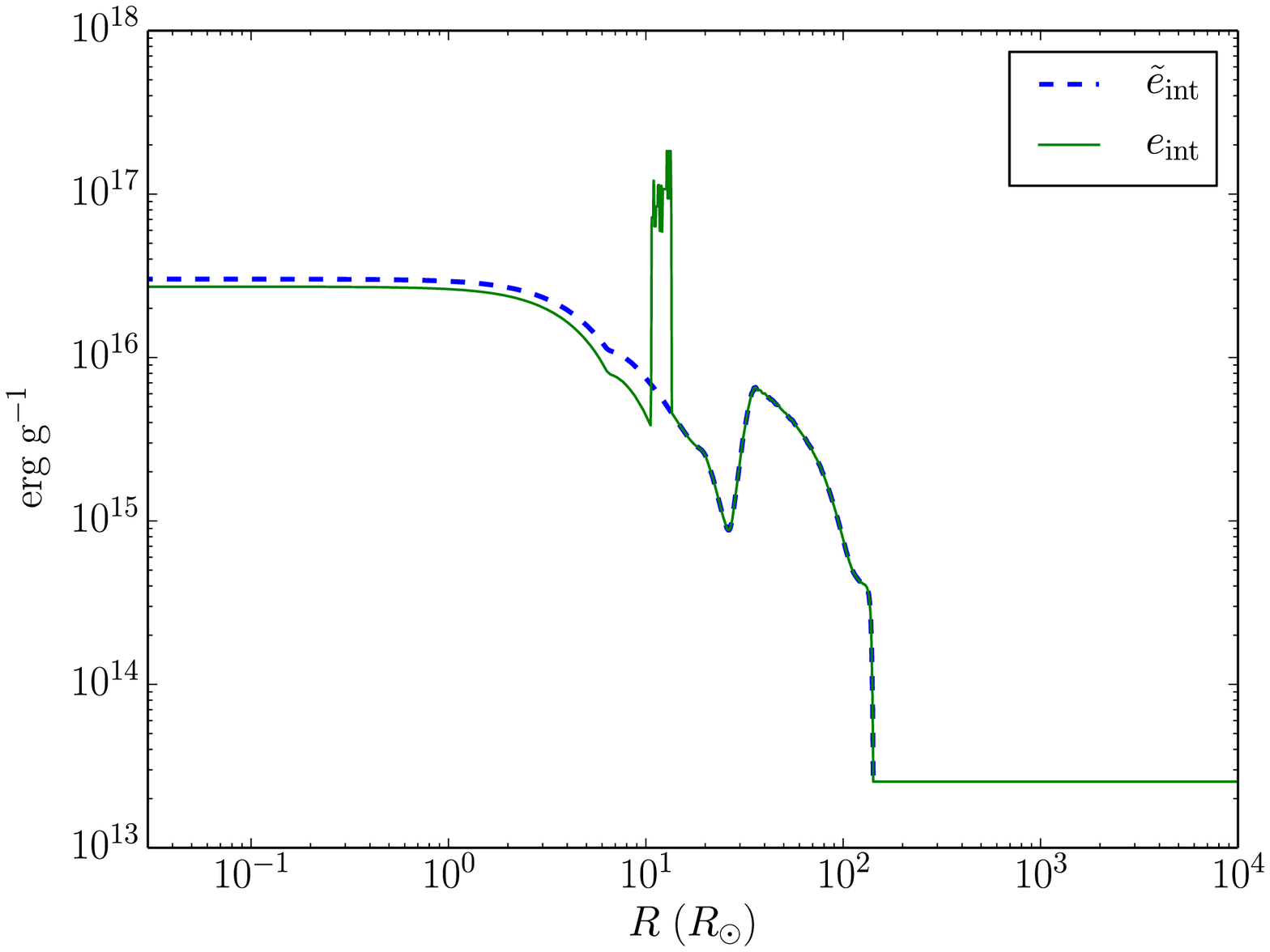}
\caption{
Left Panels:
Properties of the star after removing an outer layer of $6~M_\odot$, according to method 1, conserving energy, and adding an extension.
The specific internal (thermal) energy $\tilde{e}_{\rm{int}}$ is the energy before processing the energy conservation, and $e_{\rm{int}}$ is after the energy conservation.
Right Panels:
Properties of the star after extracting energy from the core to a layer, according to method 2, and adding an extension.
The specific internal energy $\tilde{e}_{\rm{int}}$ ($e_{\rm{int}}$) is the energy before (after) extracting the energy.
}
\label{fig:LBV2}
\end{figure*}

In one experiment we used method 1 and removed $6~M_\odot$ (5\% of
the star's mass), adjusting the energy parameters so that the ejecta
had a terminal speed of $2000 \kms$. This speed is $\sqrt 2$ the
escape velocity (not corrected for radiation pressure) of the
star at the location where the energy is deposited.
The ejected part initially had a gravitational binding energy
$E_b \approx 2 \times 10^{50}$ ergs.
The remaining nearly-hydrostatic object had a radius of only about
10.5 $R_\odot$, and the first panel in Fig. \ref{fig:LBV2} shows its properties
after the simulated eruption.

Outside the stellar radius we added a fictitious extension with negligible mass and energy,
for practical reasons noted in section \ref{sec:methods:Hydro}. The second panel in
Fig. \ref{fig:LBV2} shows a similar model using method 2. The quantitative
details may be expressed in different ways, but the basic nature
of the star in Fig. \ref{fig:LBV2} resembles a post-eruption VMS in almost any orthodox scenario.

\subsection{The Hydrodynamic Simulation}
\label{sec:methods:Hydro}
We now come to the main topic: the recovery of a VMS after a giant eruption.
After manipulating the stellar model, we import the data into the hydrodynamic code \texttt{FLASH} \citep{Fryxell2000}.
We use version 4.2.2 of the \texttt{FLASH} code. It employs an Eulerian grid with adaptive mesh refinement (AMR).
We used a 1D spherical grid with reflecting boundary conditions at the center and `diode' boundary conditions at the outer boundary,
such that gas can flow out to the guard cells but not back into the grid.
This allows mass loss from the star by the stellar wind.
We used Rosseland-mean opacity data from OPAL \citep{Iglesias1996}.
As the nuclear timescale is much longer than
the duration of the simulation, nuclear reactions are turned off in our simulation.

The stellar model was imported into \texttt{FLASH} with a low-density extension out to $r=7\times10^{14} \rm{cm} \simeq 10^4 ~R_\odot$.
This fictitious envelope merely provided space for the star to expand in the Eulerian grid.
It had $\rho \propto r^{-3}$, with negligible mass and energy.
These details had practically no effect, as that region rapidly filled with much denser material from the star.

A convection algorithm for 1D simulations, such as mixing length theory, is not incorporated.
In order to simulate convection in 1D in \texttt{FLASH}, we used a simplified prescription where
$\nabla \equiv d \ln T / d \ln P$ was limited to the adiabatic value, $\nabla \leq \nabla_{\rm{ad}}$.
This does not provide an estimate of the convective energy flux, but that quantity becomes important for our results
only in the outer regions where radiative flux dominates (section \ref{sec:results} below).

We let the star run for a few hundred years to trace its evolution.
The stability of the code was tested with a $n=3$ polytrope, and also with our VMS model with no mass removal.
In both cases the star remained in hydrostatic equilibrium and no mass was lost.

\section{Results.}
\label{sec:results}

\begin{table*}[!t]
\caption{
List of runs.
}
\begin{center}
\begin{tabular}{llllllll}
\hline \hline
Run      & Method $^1$ & Star Mass   &Mass initially removed/                     &Energy shifted/                & $R_1$ $^{3,4}$  &    Total Mass lost     \\
         &             & ($M_\odot$) &equivalent energy shifted $^2$ ($M_\odot$)  &removed ($10^{50}~\rm{erg}$)   & ($R_\odot$)     &    ($M_\odot$)         \\
\hline

1               & (1)     & 120            & 6                         &$9.11$    & 10.5          & 33                     \\
2               & (1)     & 120            & 3                         &$4.02$    & 12.3          & 6.5                    \\
3               & (1)     & 120            & 9                         &$14.6$    & 9.50          & 34                     \\

4               & (2)     & 120            & 6                         &$9.11$    &               & 11                     \\
5               & (2)     & 120            & 3                         &$4.02$    &               & 6                      \\
6               & (2)     & 120            & 9                         &$14.6$    &               & 13                     \\

7               & (2)     & 80             & 2                         &$1.79$    &               & 3.5                    \\
8               & (2)     & 80             & 4                         &$3.58$    &               & 6                      \\
9               & (2)     & 80             & 6                         &$5.30$    &               & 9                      \\
10              & (2)     & 80             & 8                         &$7.10$    &               & 14.5                   \\

\hline
\hline
\end{tabular}
\end{center}

$^1$ See text for details.

$^2$ In method 2 we shift energy from the core to outer layers. This column list the mass that this energy is able to unbind from the star.
Other sources of energy remove additional mass.

$^3$ For method 1, all layers initially outside radius $R_1$ were removed in the first step of the experiment, see text.

$^4$ Initial radii of the $120~M_\odot$ and $ 80~M_\odot$ stars were $152.8~R_\odot$ and $112.8~R_\odot$, respectively.
Their central temperatures were $4.9 \times 10^7 \K$ and $4.5 \times 10^7 \K$, respectively.

\label{Table:compareruns}
\end{table*}

\begin{figure*}
\includegraphics[trim= 0.2cm 1.7cm 1.6cm 2.8cm,clip=true,width=0.5\textwidth]{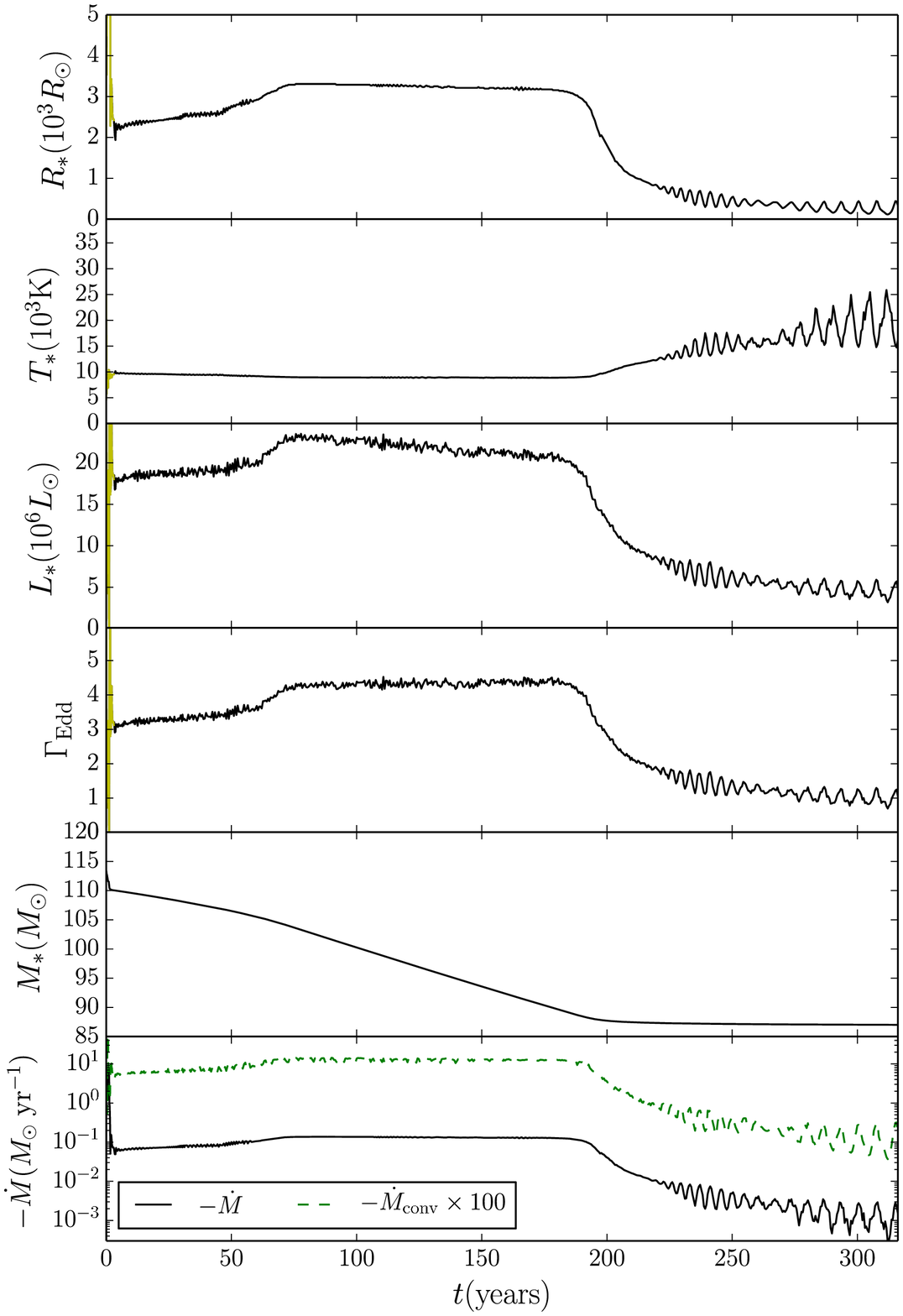}  
\includegraphics[trim= 0.2cm 1.7cm 1.6cm 2.8cm,clip=true,width=0.5\textwidth]{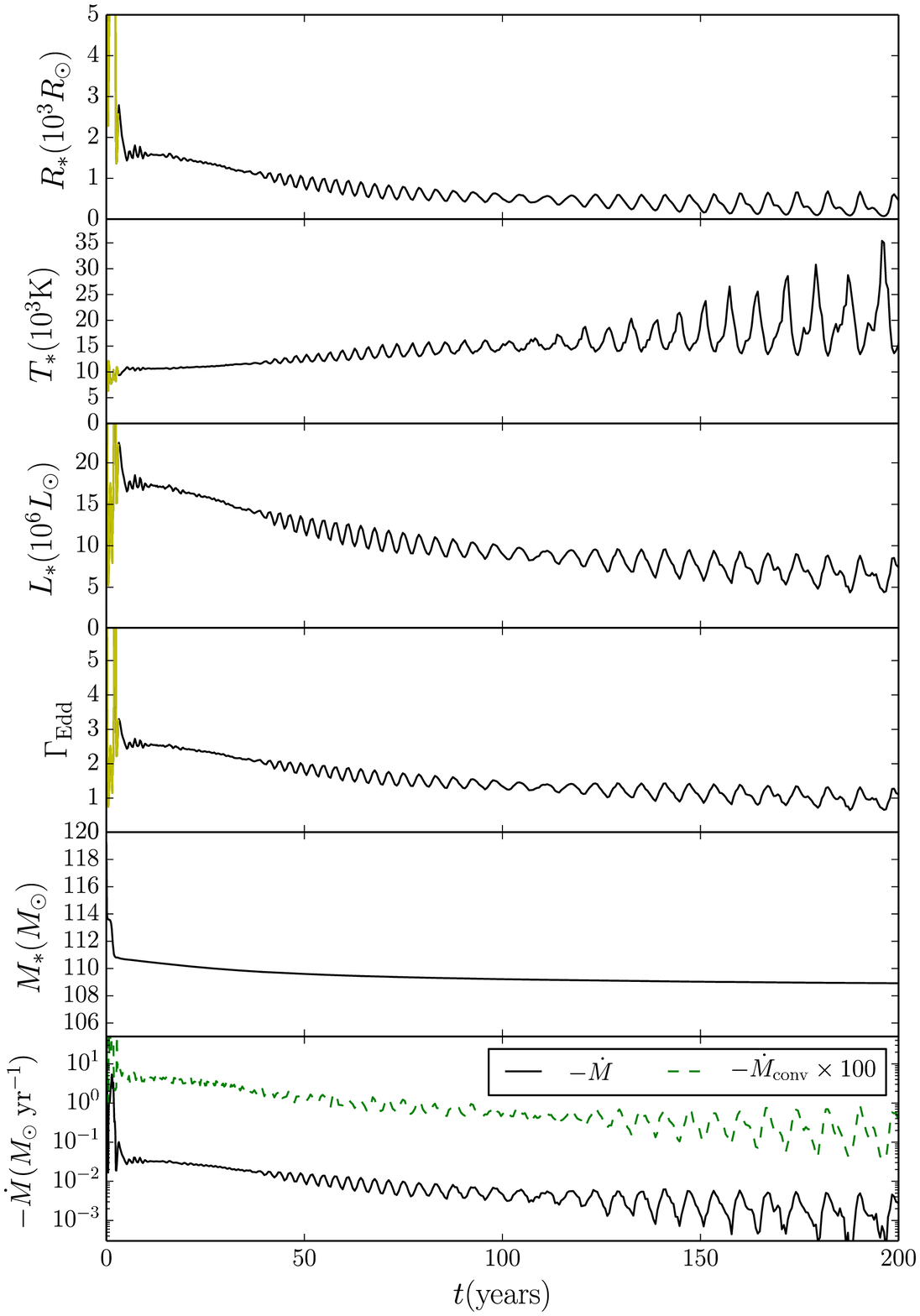}
\caption{
The properties of the star as a function of time after the eruption for run 1, in which $6~M_\odot$ is removed (left panel, method 1),
and for run 4, in which energy is extracted from the core but no mass is removed (right panel, method 2).
The panels from up to down show the stellar radius, effective temperature, luminosity, mass, Eddington ratio, and mass-loss rate.
The temperature, radius and luminosity are calculated at $\tau=3$.
As for the first $\sim 3 ~\rm{yr}$ the star is in the process of recovering from the eruption and establishing a new profile that occupies our entire grid,
results to $3 ~\rm{yr}$ are inaccurate and colored in yellow.
}
\label{fig:time_functions_multiplot}
\end{figure*}

We examine the stellar properties of the different runs at various times from the beginning of the eruption experiment ($t=0$).
starting with run 1.
Fig. \ref{fig:time_functions_multiplot} shows the properties of the star as a function of time:
stellar radius, effective temperature, luminosity, mass, Eddington ratio, mass-loss rate and convective mass-loss rate.

\begin{figure*}
\includegraphics[trim= 0.8cm 0.2cm 0.5cm 0.8cm,clip=true,width=0.5\textwidth]{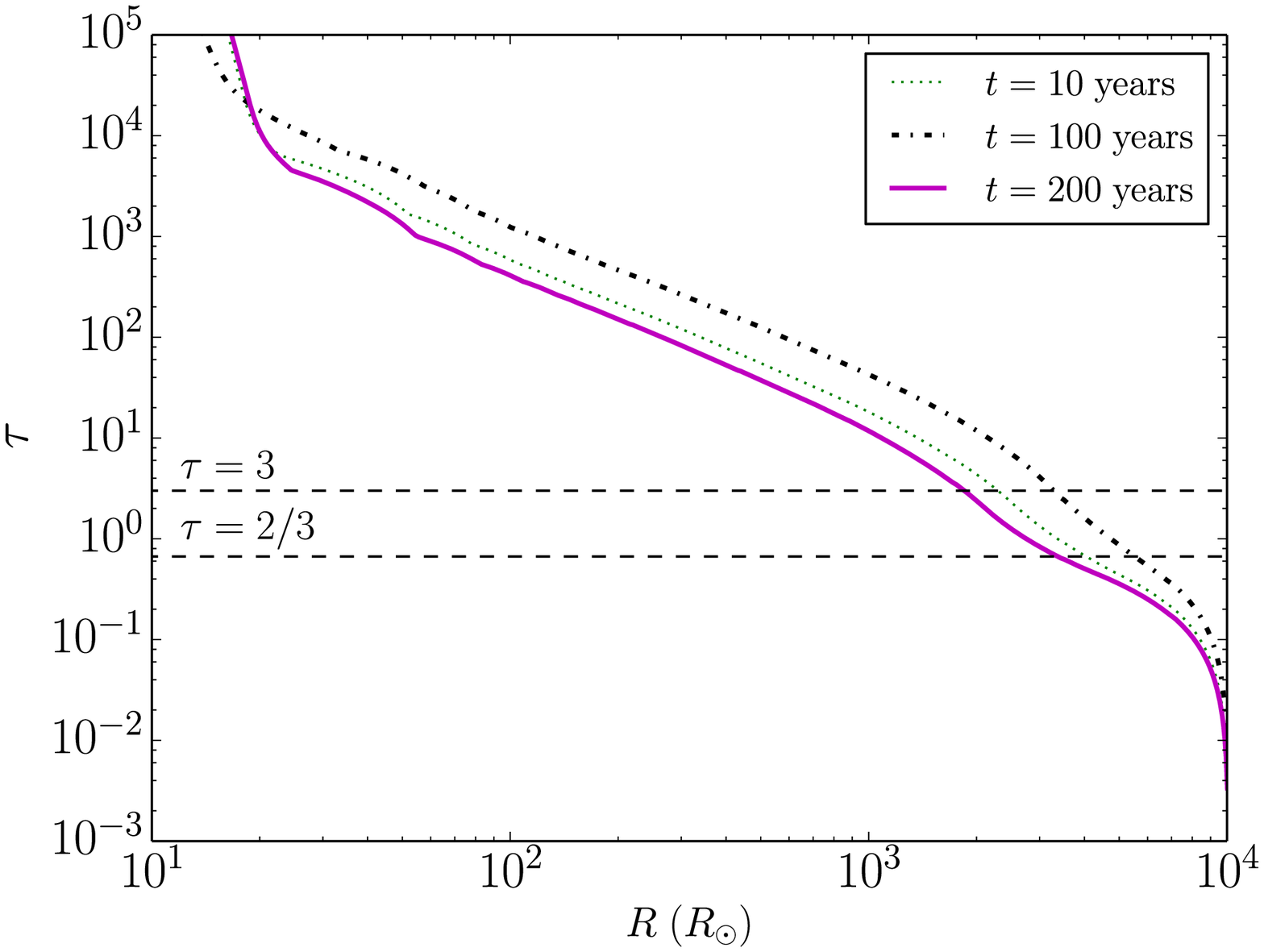}
\includegraphics[trim= 0.8cm 0.2cm 0.5cm 0.8cm,clip=true,width=0.5\textwidth]{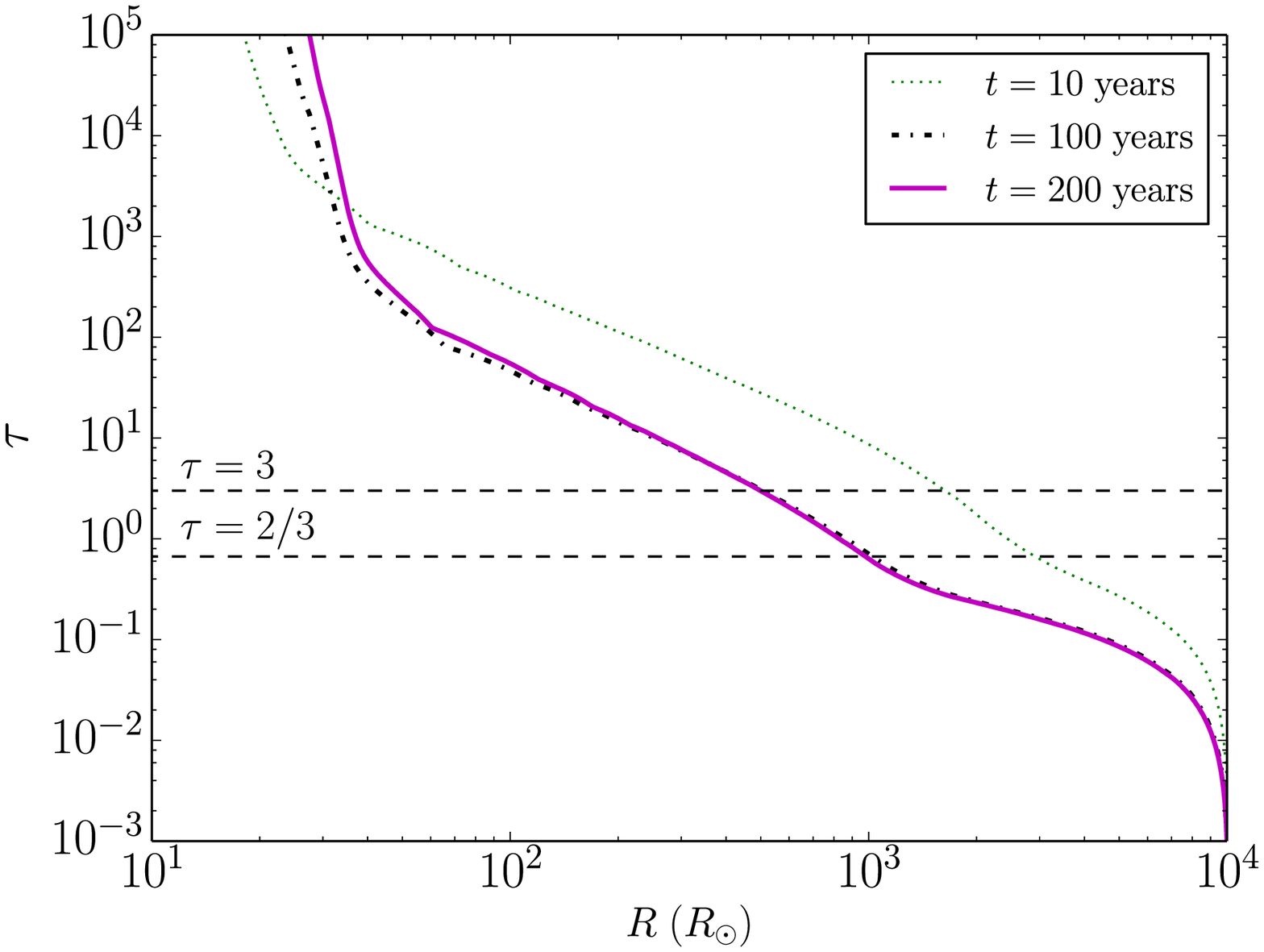}
\caption{
Optical depth of as a function of radius for different times, for run 1 (left panel), and run 4 (right panel).
Adopting $\tau=2/3$ as a measure for the location of the stellar surface gives unrealistically large results.
}
\label{fig:tau}
\end{figure*}

An observationally important question is how to define the ``apparent radius'' in a super-Eddington wind.
Fig. \ref{fig:tau}, for example, shows the exterior optical depth $\tau(R)$ in run 1 at various times.
The radius where $\tau = 2/3$, often quoted for photospheres, is very large and cool: $R \gtrsim 10^3~R_\odot$ and $T \lesssim 9000 \K$.
In fact $\tau \approx 2/3$ has no physical significance in a convex opaque wind, where most emergent photons come from $\tau > 1$.
Moreover, scattering dominates the opacity in cases of interest here.
Hence the characteristic temperature of escaping continuum radiation represents the thermalization depth, where
$\sqrt{3 \tau_{\rm{tot}} \tau_{\rm{abs}}} \approx 1$
(e.g., \citealt{HubenyMihalas2014, Davidson1987}).
In our models this occurs near $\tau_{\rm{tot}} \approx \tau_{\rm{sc}} \approx 3$,
which we adopt to define the star's apparent or observable radius $R_*$.
We find that even though the star expands to $\sim 10^3~R_\odot$ after the eruption,
its effective temperature stays above $10^4 \rm{K}$, on the blue side of the HD limit on the HR diagram.

\begin{figure*}
\includegraphics[trim= 0.0cm 0.5cm 1.2cm 1.2cm,clip=true,width=0.5\textwidth]{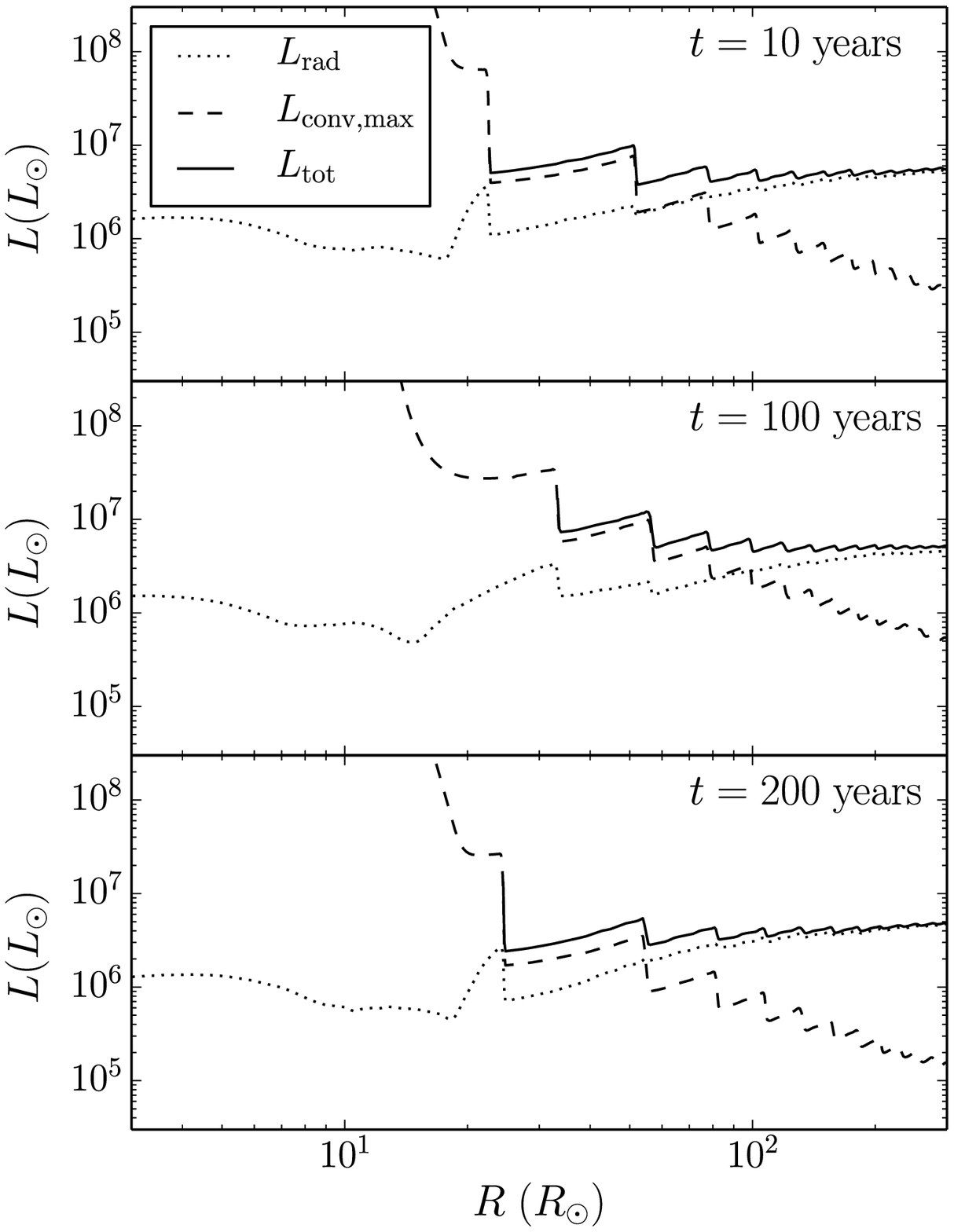}
\includegraphics[trim= 0.0cm 0.5cm 1.2cm 1.2cm,clip=true,width=0.5\textwidth]{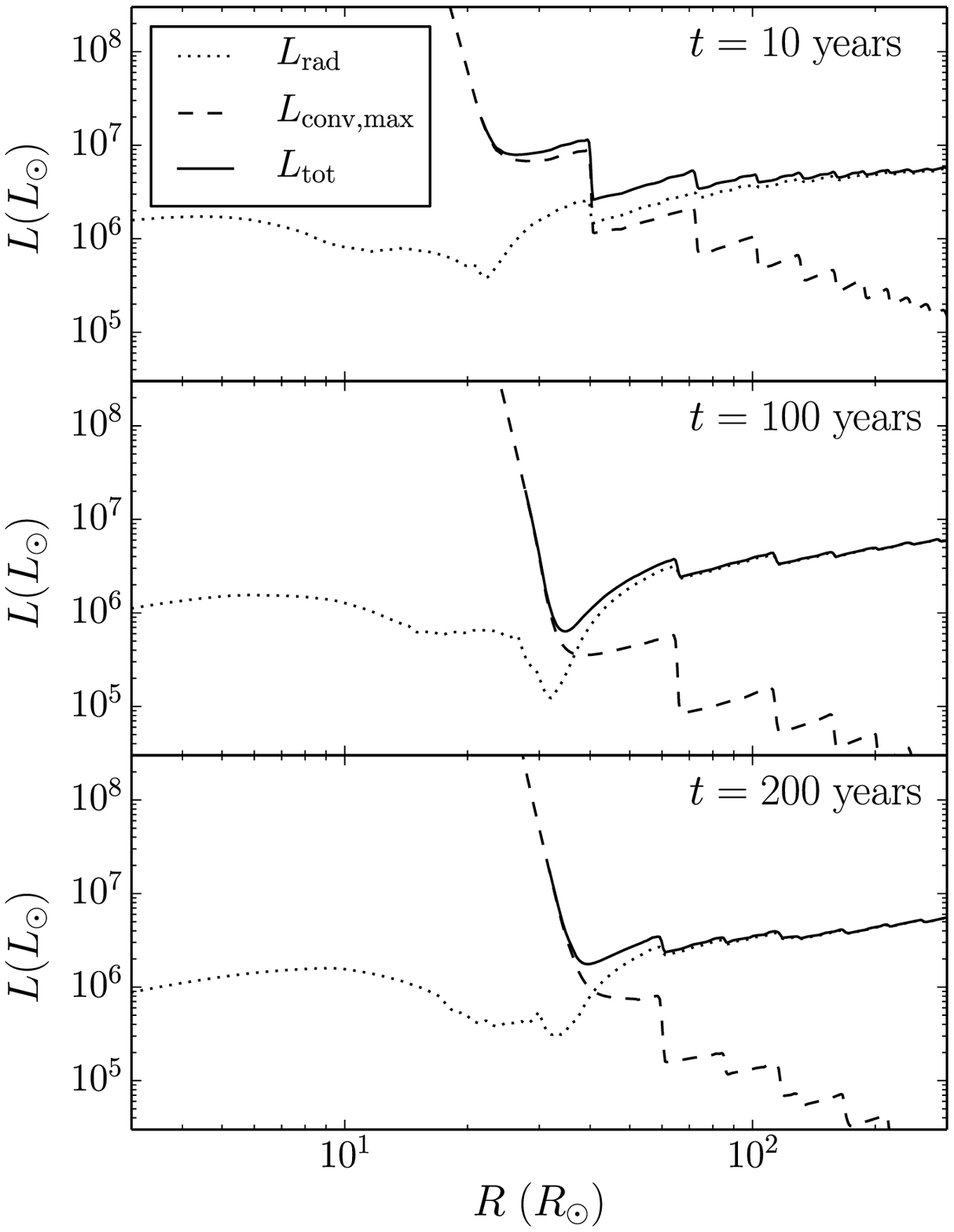}
\caption{
The radiative, convective and total local luminosity, at different times, for run 1 (left panel), and run 4 (right panel).
The dependence on $r$ is entirely a consequence of the star not being in thermal equilibrium;
it is not related to the nuclear burning rate which has a much longer timescale.
At outer radii the radiative luminosity dominates.
The radiative luminosity is calculated using the diffusion approximation.
Note that at large radii $L_{\rm{rad}}$ (and hence $L_{\rm{tot}}$) is overestimated as due to the low density the mean free path of the photons is
larger than their remaining travel distance.
The convective luminosity shown here is only a maximum value.
}
\label{fig:Luminosities}
\end{figure*}

Another question is how to define the luminosity when the star is far from equilibrium.
Fig. \ref {fig:Luminosities} shows the radiative luminosity $L_{\rm{rad}}$, maximum convective luminosity $L_{\rm{conv,max}})$,
and the total luminosity $L_{\rm{tot}}$, at different times for runs 1 and 4 (discussed below).
The local luminosity depends on $r$ because the star is not in thermal equilibrium.
The radiative luminosity is calculated using the diffusion approximation
\begin{equation}
L_{\rm{rad}}(r) \approx - \frac{16 \pi r^2 a c T^3} {3 \kappa \rho}  \frac{\partial T}{\partial r}
\label{eq:Lrad_r}
\end{equation}
where $r$ is the radius, $T$ is the temperature, $\kappa$ is the opacity, and $a$ is the radiation constant.
At large radii, when approaching the stellar surface, and in the wind,
the estimate of $L_{\rm{rad}}$ based on the diffusion approximation in eq. (\ref{eq:Lrad_r}) becomes invalid.
The reason is that the because of the low density, the mean free path of photons becomes
comparable with, and finally larger than, the remaining travel distance to the stellar surface.
This results in a large overestimate of $L_{\rm{rad}}$.
We therefore plot it only up to $r=100~R_\odot$.
The region at a few $\times 10~R_\odot$ shows strong decline in $L_{\rm{rad}}$.
We will return to this detail later.

Only a {\it maximum} convective luminosity can be calculated here:
\begin{equation}
L_{\rm{conv,max}}(r) \lesssim \pi r^2 \rho v_s^3
\label{eq:Lconv_r}
\end{equation}
where $v_s$ is the local speed of sound. The true value is presumably smaller.
Fig. \ref {fig:Luminosities} shows that the luminosity at $r \gtrsim 100~R_\odot$ is dominated by radiation.
The Eddington ratio is calculated using the Thomson scattering opacity because the absorption fraction is small.
\begin{figure}
\includegraphics[trim= 0.2cm 1.7cm 1.6cm 2.3cm,clip=true,width=0.5\textwidth]{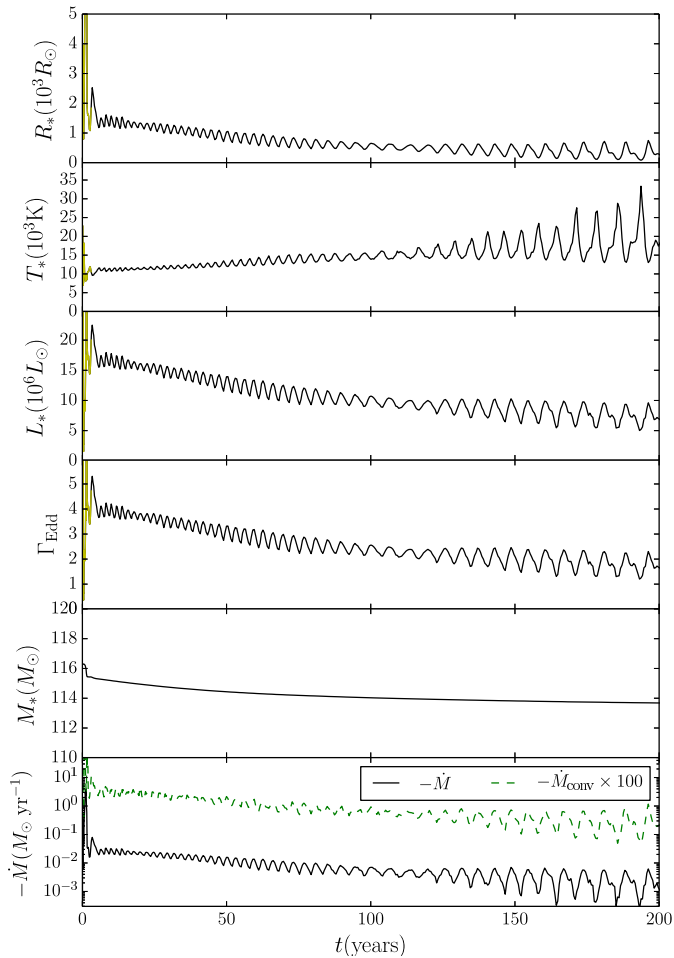}  
\caption{
The properties of the star as a function of time after the eruption for run 2, in which $3~M_\odot$ is removed using method 1.
Panels are the same as in Fig. \ref{fig:time_functions_multiplot}.
The history of $\eta$ Car resembles a run in between this run and run 1.
}
\label{fig:time_functions_multiplot_run2}
\end{figure}

Table \ref{Table:compareruns} shows the mass-loss results for ten calculative runs.
The time dependence of run 1, for example, is portrayed in Fig. \ref{fig:time_functions_multiplot}.
This model {\it continued to eject material at a high rate long after
the initial event,\/} and ultimately lost more than three times
as much mass as the amount that we artificially removed to initiate
the process.
For several years the star approached a sort of quasi-equilibrium and expanded into our coordinate grid, and then,
during the interval $10 \rm{yr} \lesssim t \lesssim 200 \rm{yr}$, it had a
mass-loss rate of the order of $0.1 ~M_\odot~\rm{yr^{-1}}$.
This qualitatively resembles the case of $\eta$ Car, but was more extreme and more protracted.
There are strong hints that $\eta$ Car had a rate $\dot{M} \gtrsim 10^{-2} ~M_\odot~\rm{yr^{-1}}$ a century ago
\citep{Humphreys2008}, and by the year 2000 this had declined to
roughly $10^{-3} ~M_\odot~\rm{yr^{-1}}$ (\citealt{DavidsonHumphreys1997}; \citealt{Hillier2001}).
Thus a model intermediate between our runs 1 and 2 would somewhat resemble $\eta$ Car's history in the past 160 years.
As explained below, the underlying cause for mass-loss both in that stage and
later is a $\kappa$-mechanism induced pulsation.
As $\dot{M}$ declined in run 1 after $t \sim 200$ yr, most of the quantities in
Fig. \ref{fig:time_functions_multiplot} showed obvious outward-moving waves.

VMSs have regions with strong convection.
In the outer layers of the star, where convection becomes inefficient due to decline in $\rho v_s^3$ (where $v_s$ is the speed of sound), the
radiative Eddington factor
\begin{equation}
\Gamma_{\rm{rad}}=\frac{\kappa L_{\rm{rad}}}{4 \pi r^2 g c}
\label{eq:Gamma_rad}
\end{equation}
(where $L_{\rm{rad}}$ is the radiative luminosity, and $g=GM/R$)
becomes larger than 1.
The radius where this occurs is called the critical radius $R_c$, and the density at this location is the critical density $\rho_c$.
If we put the sonic radius for a wind outflow at this location $v_{s,c}$ we can derive a convective mass-loss rate
\begin{equation}
\dot{M}_{\rm{conv}} \approx 4 \pi R_c^2 \rho_c v_{s,c}.
\label{eq:mdot_conv}
\end{equation}
The sixth row on the left panel of Fig. \ref{fig:time_functions_multiplot} shows $\dot{M}_{\rm{conv}}$ for run 1 together with the actual mass-loss rate.
It can be seen that the actual mass-loss rate is very close to the convective mass-loss rate.

The reason for the decrease in mass-loss rate at later times is that as mass is lost the profile of the star changes, and layers with steeper
density gradient are exposed.
These layers have a higher gravitational binding energy, and therefore are more difficult to remove.

\begin{figure*}
\begin{center}
\includegraphics[trim= 0.0cm 0.1cm 0.2cm 1.9cm,clip=true,width=0.80\textwidth]{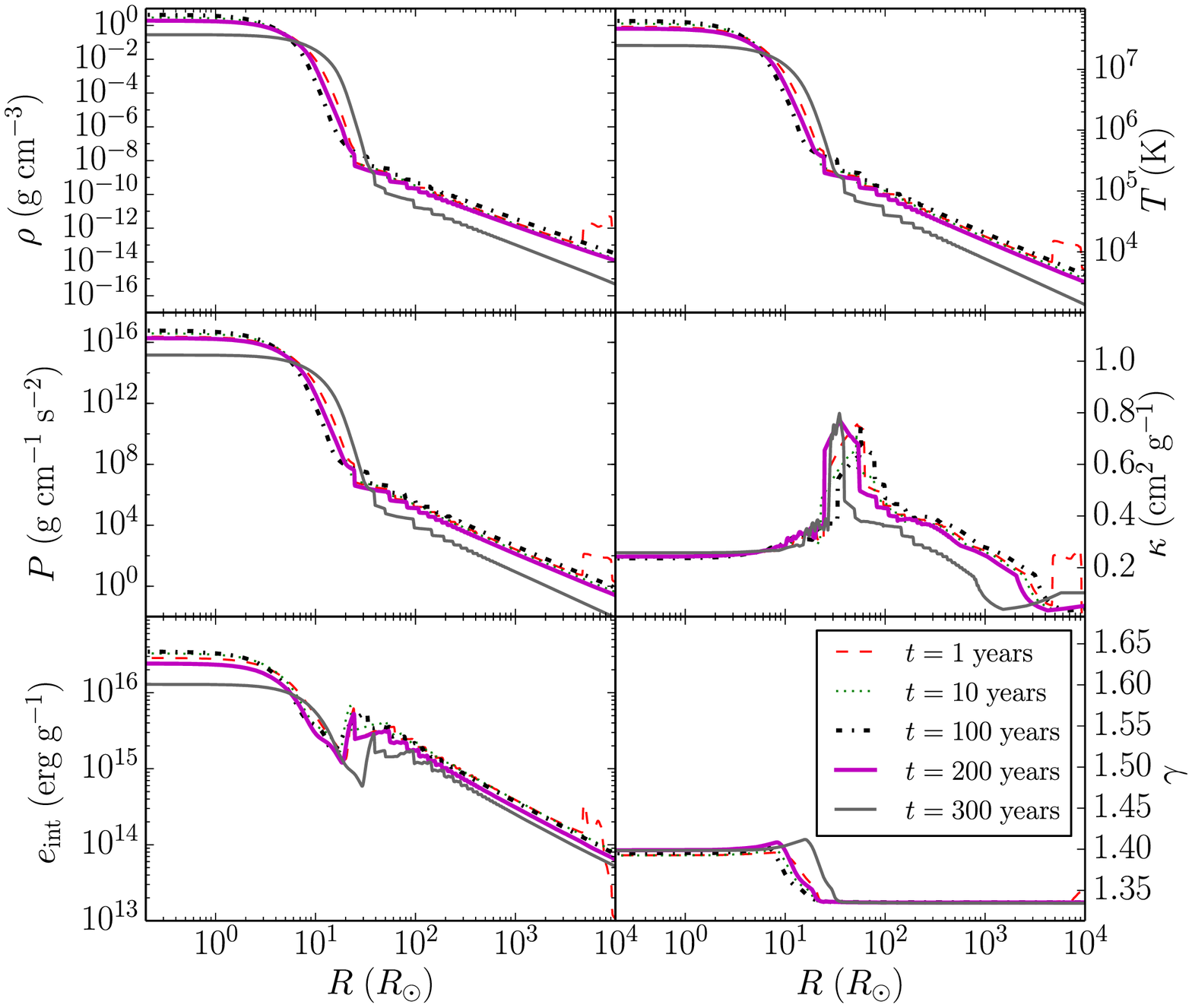}   
\includegraphics[trim= 0.0cm 0.1cm 0.2cm 1.9cm,clip=true,width=0.80\textwidth]{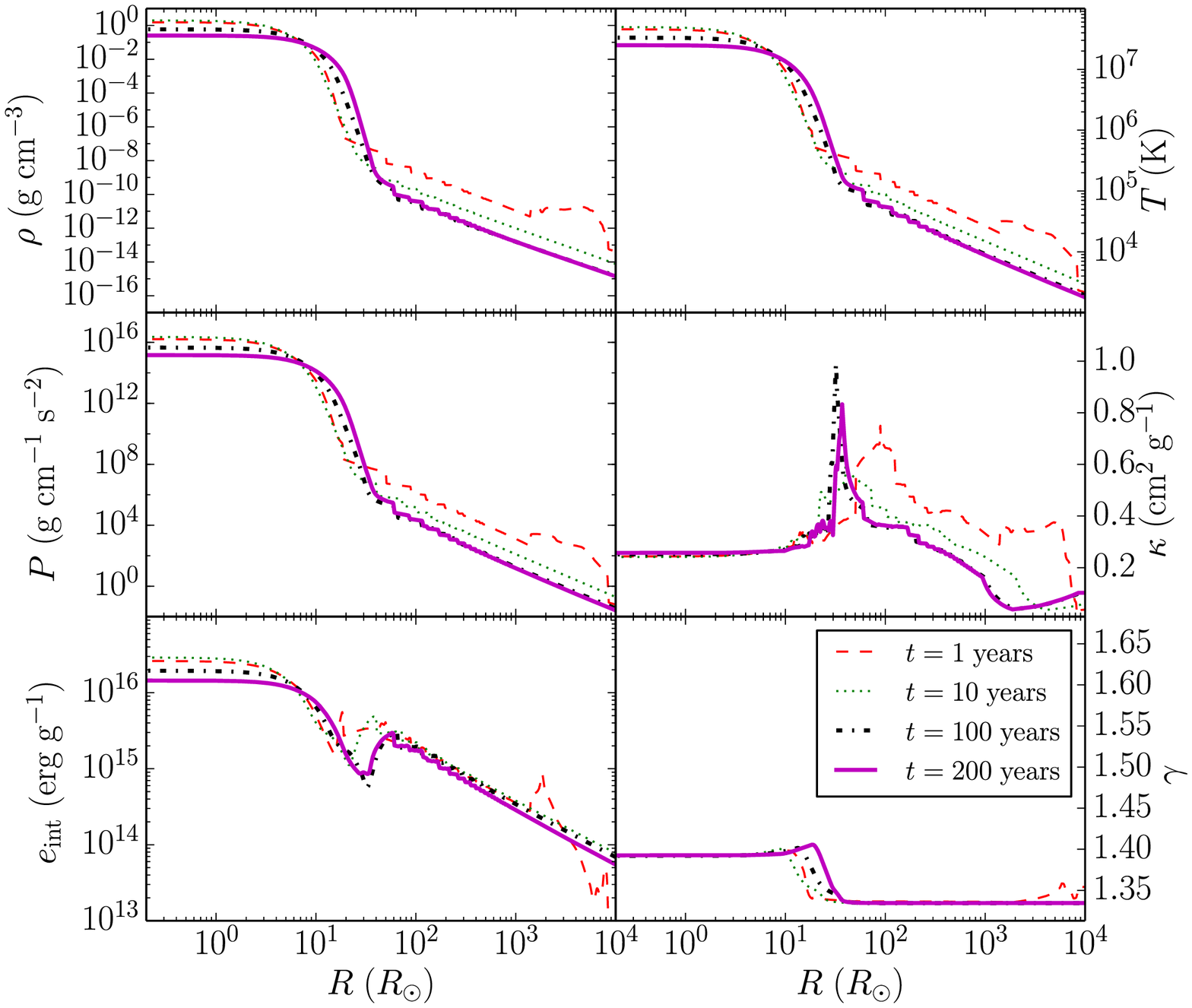}
\end{center}
\caption{
The properties of the star as a function of radius (the radius includes the extension),
at different times after the eruption, for run 1 (top panel), and run 4 (bottom panel).
At $t=1~\rm{year}$ the original extension is still visible at $r > \rm{few} \times 10^3~R_\odot$.
}
\label{fig:multiplot}
\end{figure*}

In Fig. \ref{fig:multiplot} we show the properties of the star as a function of radius (including the extension),
for different times after the eruption.
Note the following phases:
\begin{enumerate}
\item At early times the original low density - low pressure extension still exists.
It can be seen as a very inconspicuous feature at $r \simeq 5000 ~R_{\odot}$ in the line representing $t = 1 ~\rm{yr}$ in upper panel of the Fig. \ref{fig:multiplot}.
It takes the density profile (of the star and its wind) $\sim 2.5 ~\rm{yr}$
to expand to the edge of the grid.
As mentioned, we experimented with different extensions, and found that the original extension profile does not affect the results, if the mass
in the extension is much smaller then the stellar mass.
\item The core of the star was initially compressed as a result of the eruption.
After $\sim 1.5 ~\rm{yr}$ it returned to its original density.
\item After $\sim 2.5 ~\rm{yr}$ the star develops a structure that conserves its shape as time goes by.
The profiles do change as a result of an ongoing mass-loss, but the general shape remains roughly the same.
\end{enumerate}
The behavior of the star at $t \lesssim 10 ~\rm{yr}$ might not be very meaningful as it depends on the details
of the method we use. That being said, for the two methods we tried we do however get similar results for that initial time after the eruption.

By examining the regions of high opacity, we can interpret the decrease in mass loss after $t \sim 200$ yr.
The strong opacity peak is identified as the ``Iron Bump'', located at $\sim 200\,000 \rm{K}$.
Before $t \sim 200$ yr the location of the peak is in relatively low density region, such that the pulsations
that are strongest at that region, can push the gas outwards and result in the high mass loss.
As mass is depleted from the star, its structure changes.
As it approaches $t \sim 200$ yr, the opacity peak location coincides with the high slope of the density towards the inner layers of the star.
These layers are more massive and cannot be easily pushed out by the pulsations.
The energy of the pulsations still travels out, removing the low density outer layers of the star.
Gradually over a few decades, the density in the outer layers becomes smaller, and the mass-loss rate decreases.
As the density decreases so is the optical depth, therefore the location of $\tau=3$ by which we determine the radius moves inwards.
During that time the lower density in outer layers causes the density profile of the star to expand,
such that the density gradient becomes smaller, and the density in the center decreases.

\begin{figure*}
\includegraphics[width=0.5\textwidth]{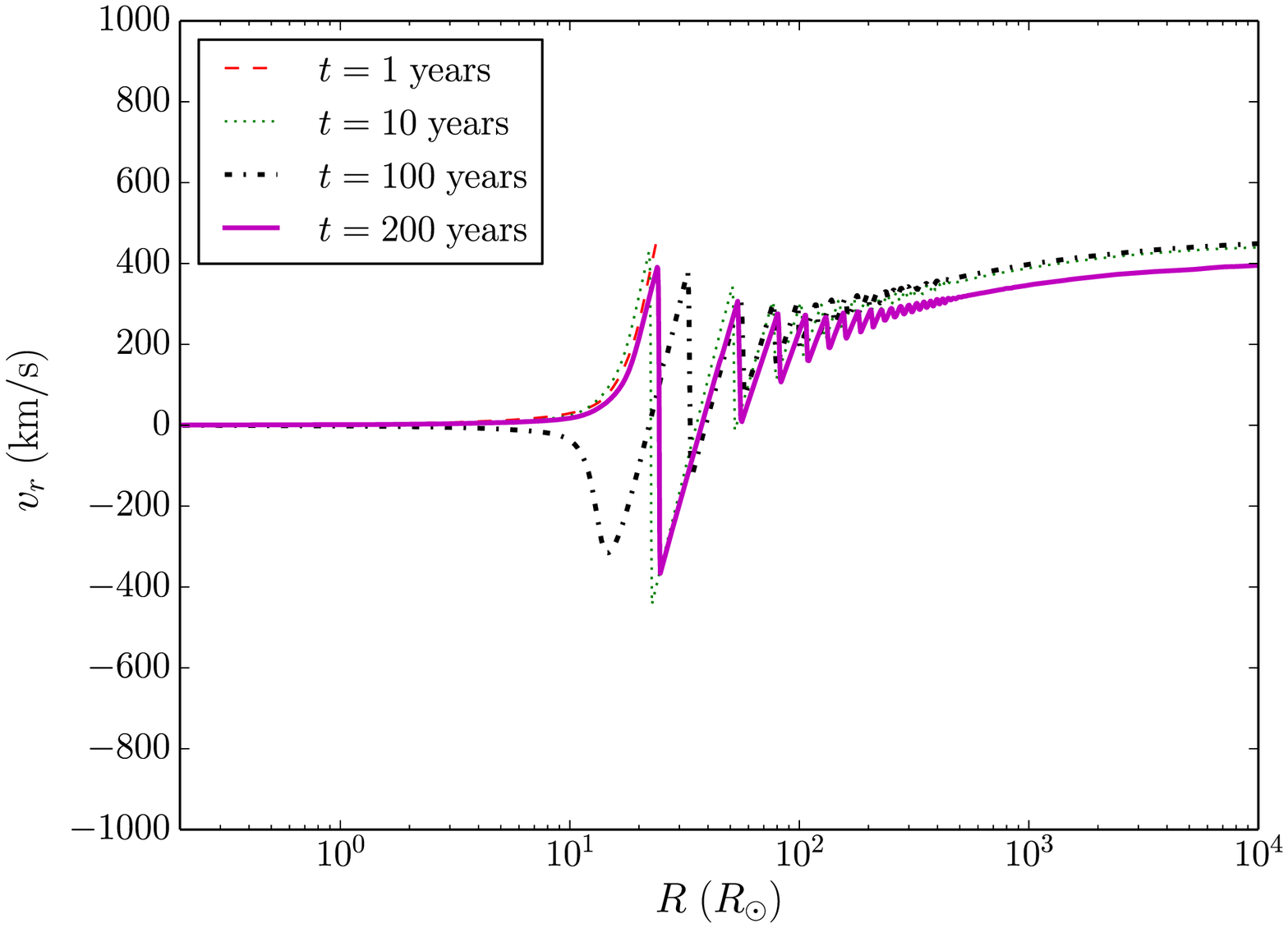}
\includegraphics[width=0.5\textwidth]{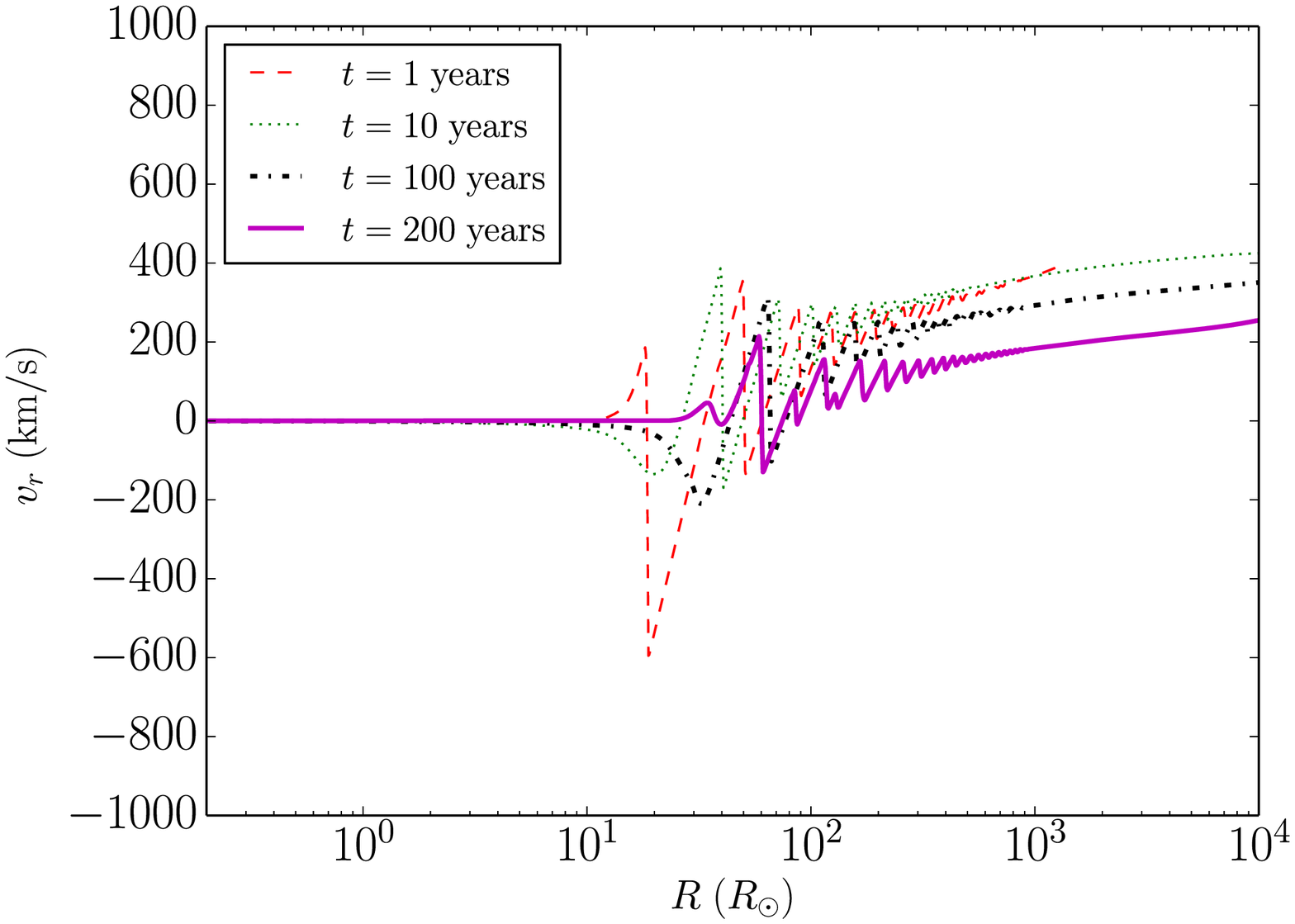}
\caption{
The radial velocity profile of the star at different times after the eruption, for run 1 (left panel), and run 4 (right panel).
We do not plot the velocity at the location of the original extension for times where the original extension still exists.
The lower panel zooms in on the velocities at late times, showing the region that strongly fluctuates between positive (outward) and negative (inward) values,
and the outer region where the wind accelerates.
}
\label{fig:velx}
\end{figure*}

The left panel of Fig. \ref{fig:velx} shows the radial velocity profiles for run 1 (left panel), and run 4 (right panel)
Ripples in the velocity profile (and also in the pressure profile -- see third panel of Fig. \ref{fig:multiplot}) started developing.
Each layer expands into low density and pressure, and then the following layer expands into
a denser region and with a higher pressure, and so on.
The exact shape of the ripples is a result of the stellar density and pressure profile.
In a 3D model, presumably the ripples shown
in our figures become much less coherent when averaged over
latitude and longitude -- i.e., waves propagate in various directions
with random elements not present in our radial 1D calculations.
The cumulative effect, however, would be an outward mass and energy flow.

\begin{figure*}
\includegraphics[width=0.5\textwidth]{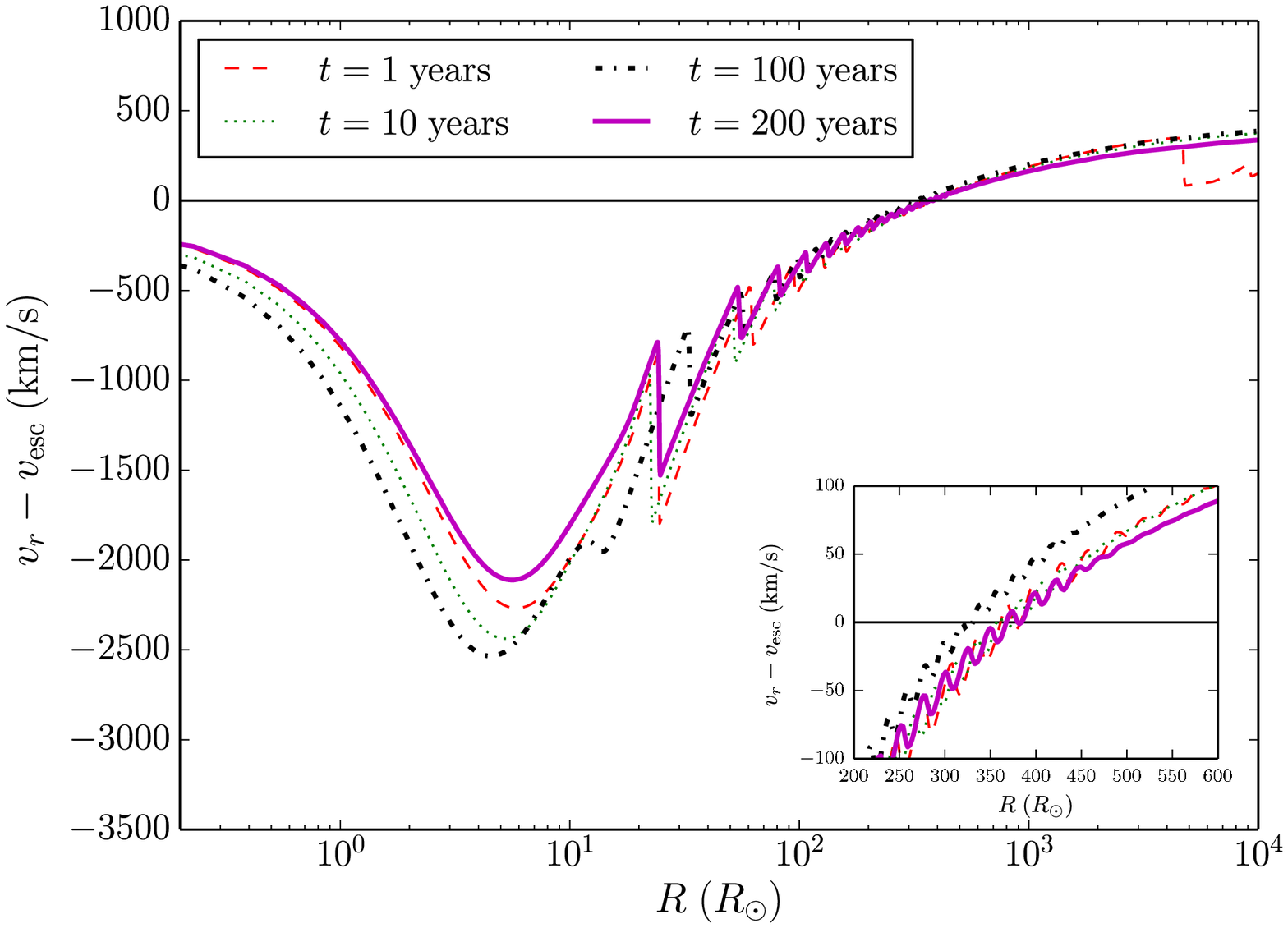}
\includegraphics[width=0.5\textwidth]{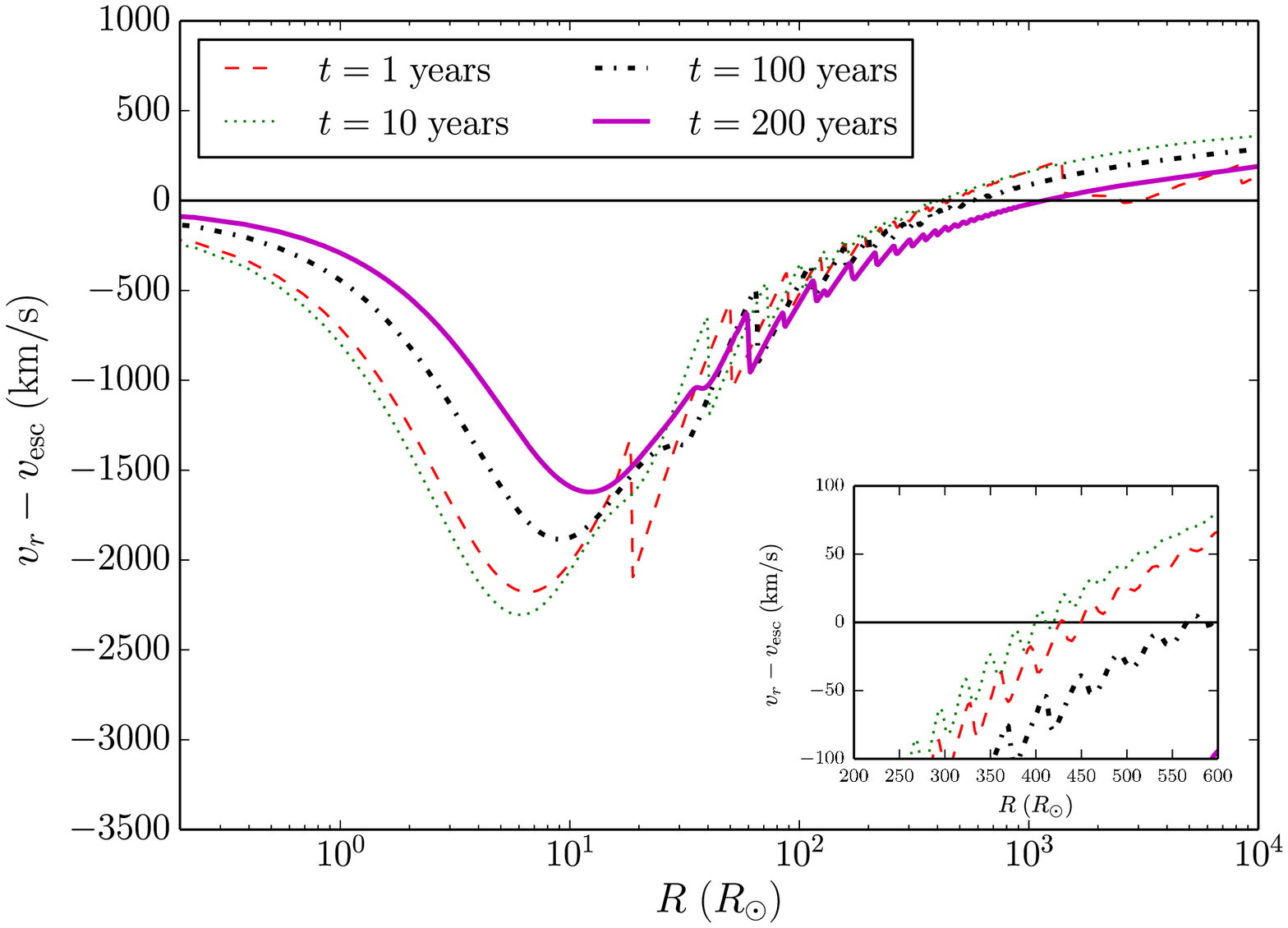}
\caption{
The radial velocity minus the escape velocity, at different times after the eruption, for run 1 (left panel), and run 4 (right panel).
}
\label{fig:velx-vesc}
\end{figure*}

At later times a more or less similar velocity profile is observed,
with the velocity at the outer edge of the grid $v_{\rm{out}} \simeq 430 \rm{km~s^{-1}}$.
This is comparable to the terminal velocities observed in VMSs.
The velocity profile has a region where the velocity fluctuates between positive (outward) and negative (inward) values,
as can be seen in Fig. \ref{fig:velx}.
This is the region where we found pulsations (see below).
Further out the star develops a wind that causes it to lose mass.
Fig. \ref{fig:velx-vesc} shows the radial velocity after subtracting the escape velocity.
This is the radius where the gas escapes from the star.

\begin{figure*}
\includegraphics[trim= 0.4cm 0.4cm 0.6cm 0.8cm,clip=true,width=0.5\textwidth]{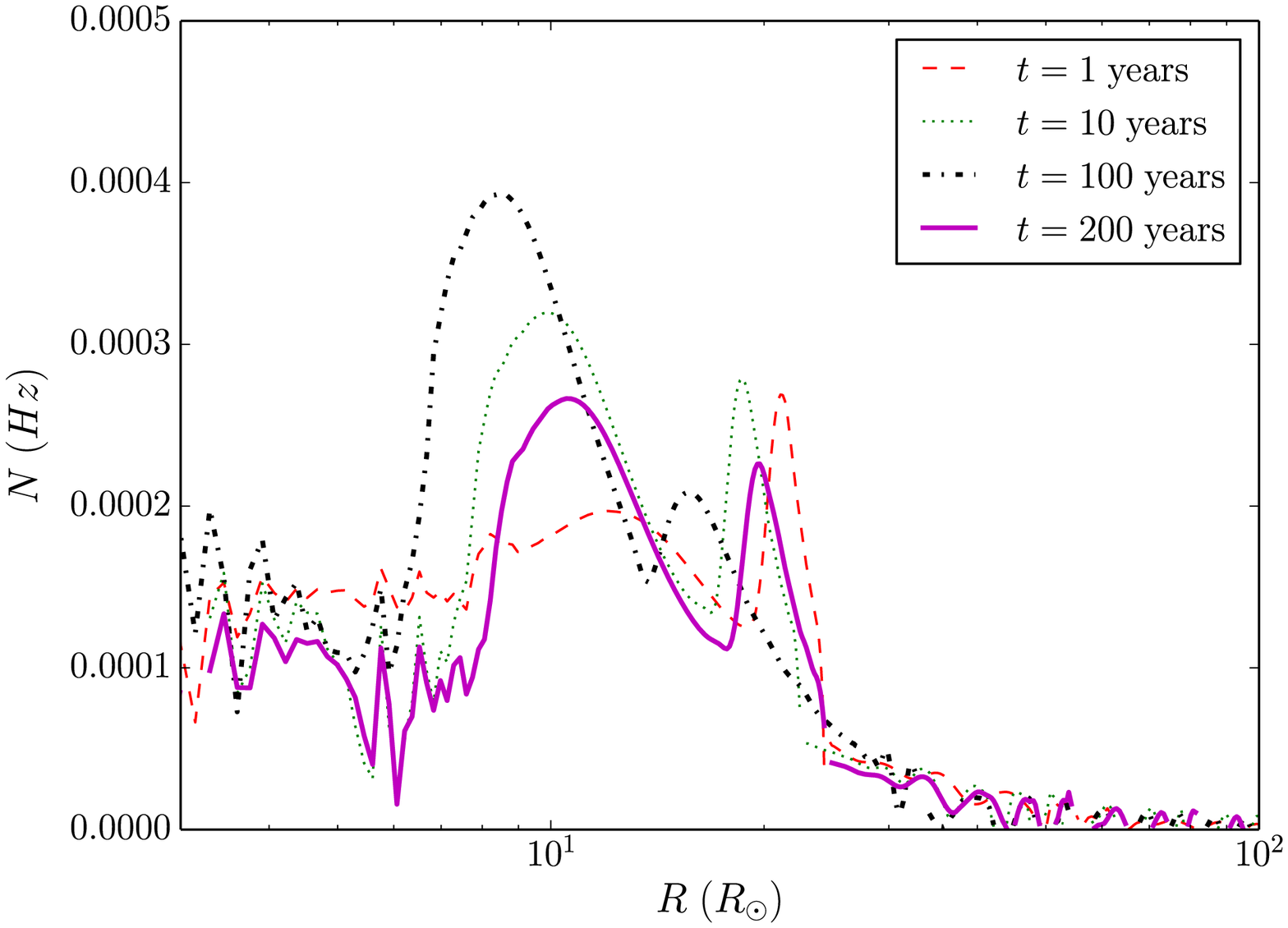}
\includegraphics[trim= 0.4cm 0.4cm 0.6cm 0.8cm,clip=true,width=0.5\textwidth]{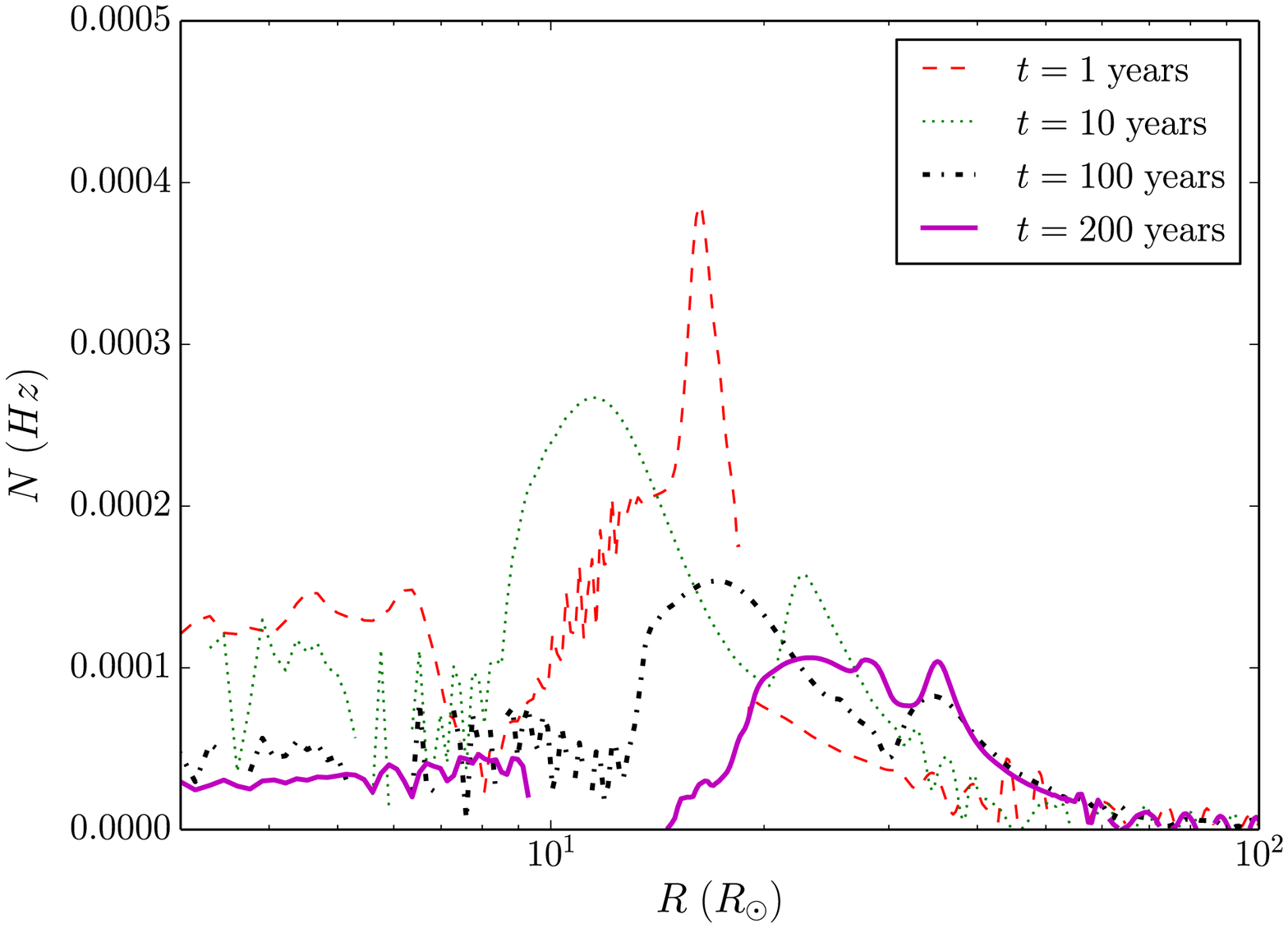}
\caption{
The Brunt-V\"{a}is\"{a}l\"{a} frequency of the star as a function of radius, for different times, for run 1 (left panel), and run 4 (right panel).
}
\label{fig:BV}
\end{figure*}

We calculate the Brunt-V\"{a}is\"{a}l\"{a} frequency of the star $N_{\rm{BV}}$ (the frequency for adiabatic pulsations)
\begin{equation}
\frac{1}{g}N_{\rm{BV}}= \frac{1}{\gamma} \frac{d\ln P}{dr}  -  \frac{d\ln \rho}{dr}
\label{eq:N_BV}
\end{equation}
as a function of radius, for different times (Fig. \ref{fig:BV}),
where $g$ is the gravitational acceleration and $\gamma$ is the adiabatic index.
There is a strong peak in the same layers where the pulsation occurs.
As time moves on, the peak moves slightly outwards, indicating that the strong pulsation layer is propagating as mass is lost from the star
and the density becomes smaller.

\begin{figure*}
\includegraphics[trim= 0.3cm 0.5cm 1.6cm 2.2cm,clip=true,width=0.5\textwidth]{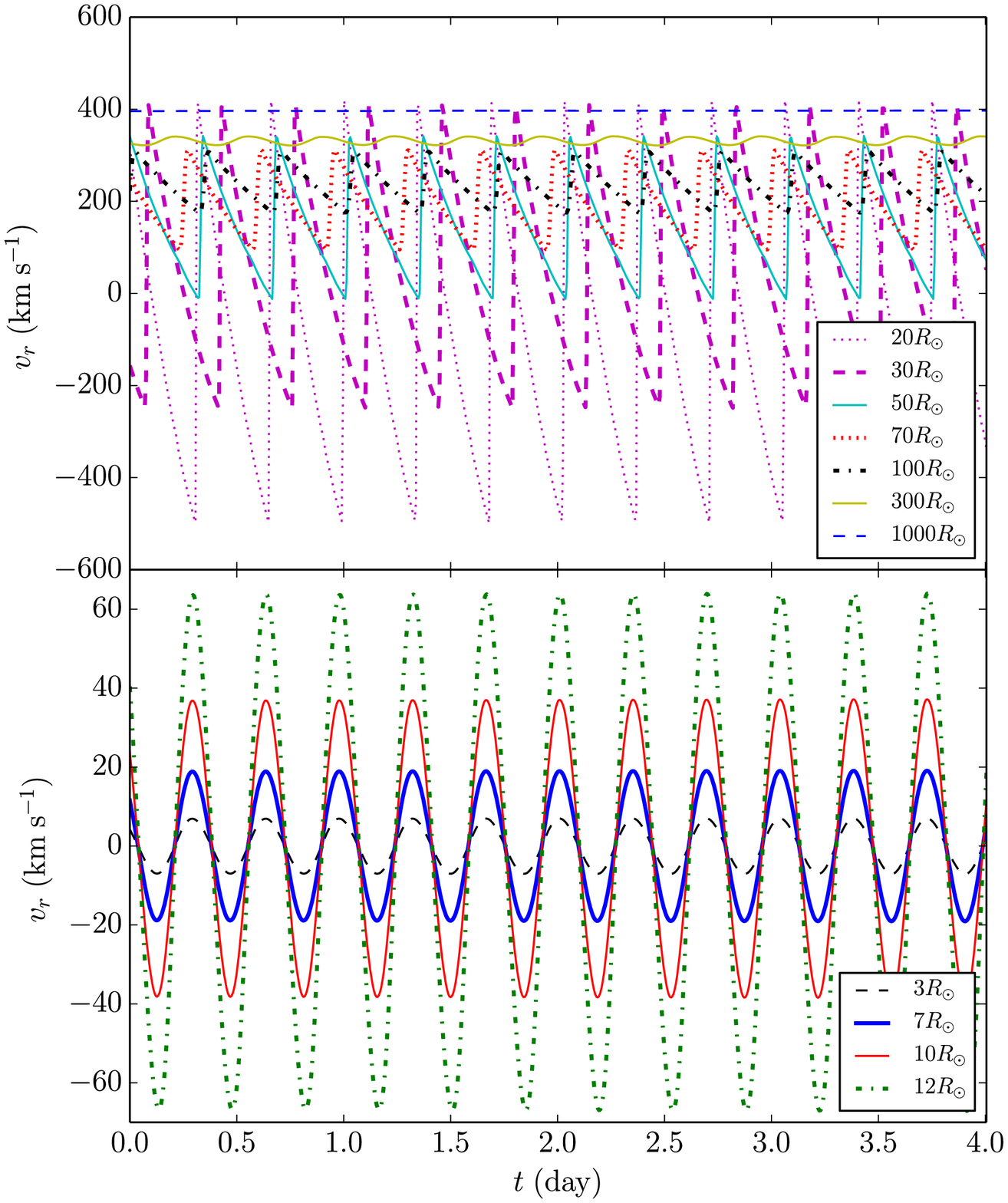}
\includegraphics[trim= 0.3cm 0.5cm 1.6cm 2.2cm,clip=true,width=0.5\textwidth]{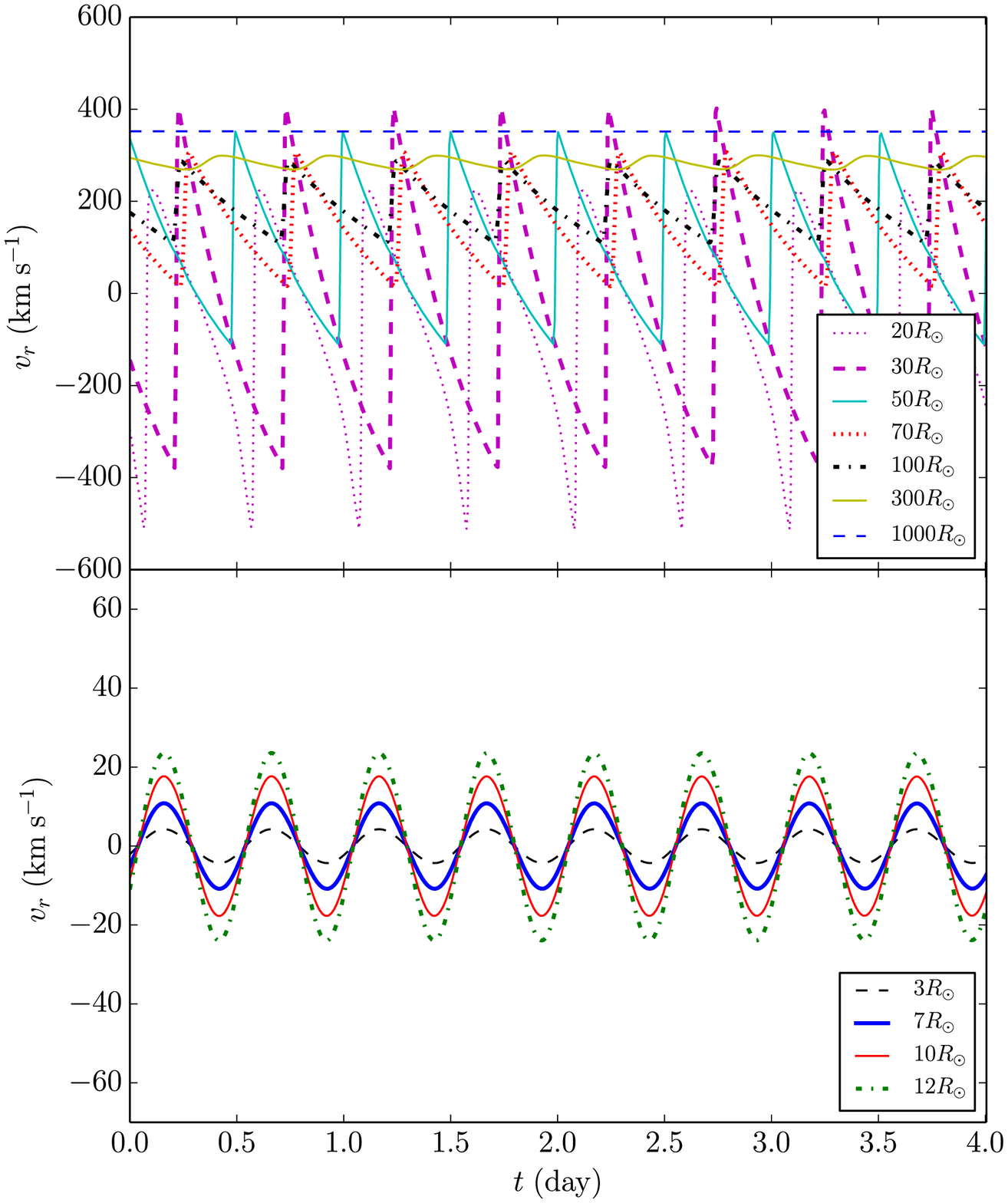}
\caption{
The radial velocity at different radii as a function of time, after $t=10^9 \rm{s}$, for run 1 (left panel), and run 4 (right panel).
Positive velocities indicate flow in the positive radial direction.
}
\label{fig:velx_t}
\end{figure*}

\begin{figure*}
\includegraphics[trim= 0.3cm 0.15cm 0.4cm 0.35cm,clip=true,width=0.5\textwidth]{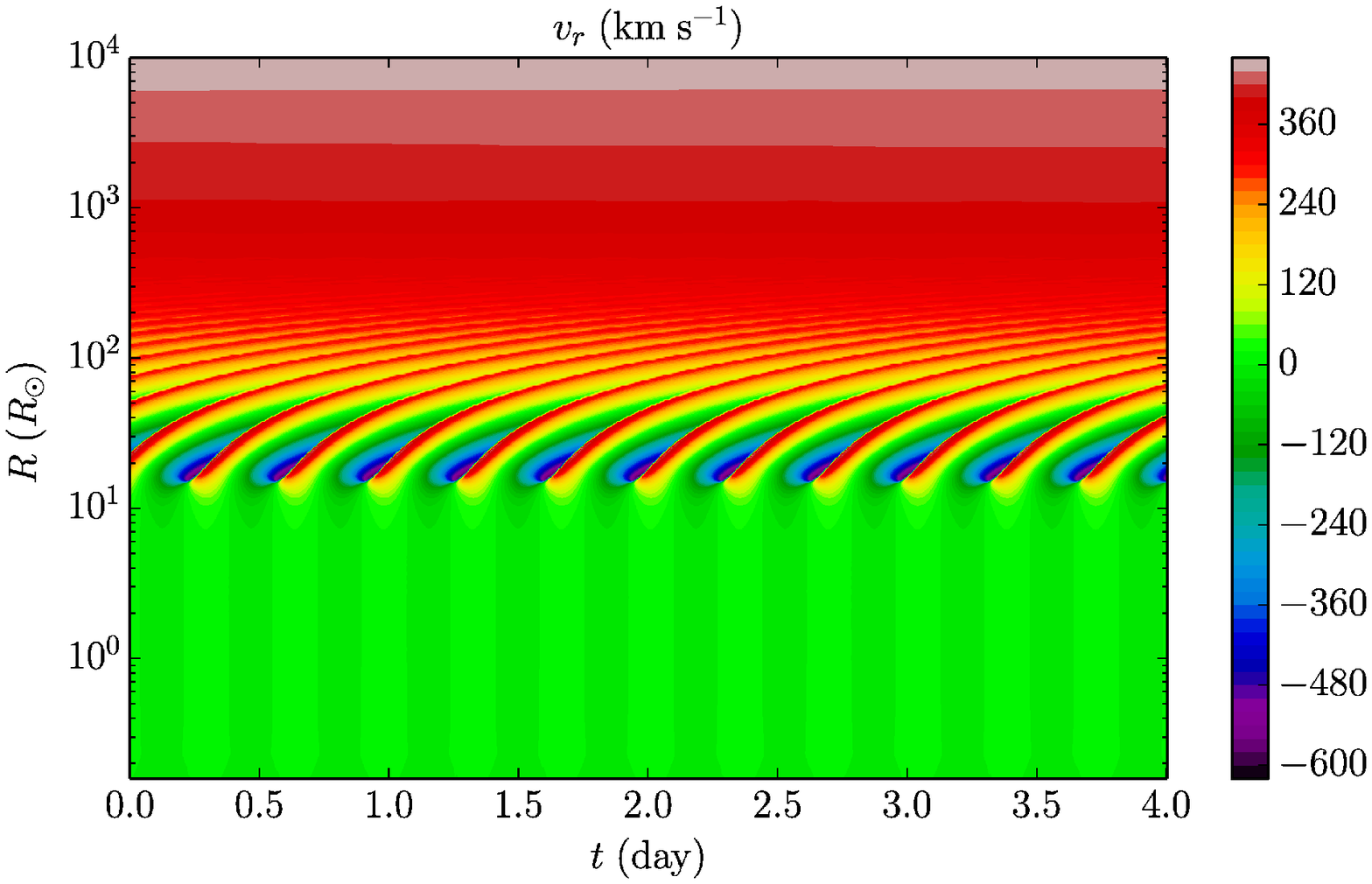}  
\includegraphics[trim= 0.3cm 0.15cm 0.4cm 0.35cm,clip=true,width=0.5\textwidth]{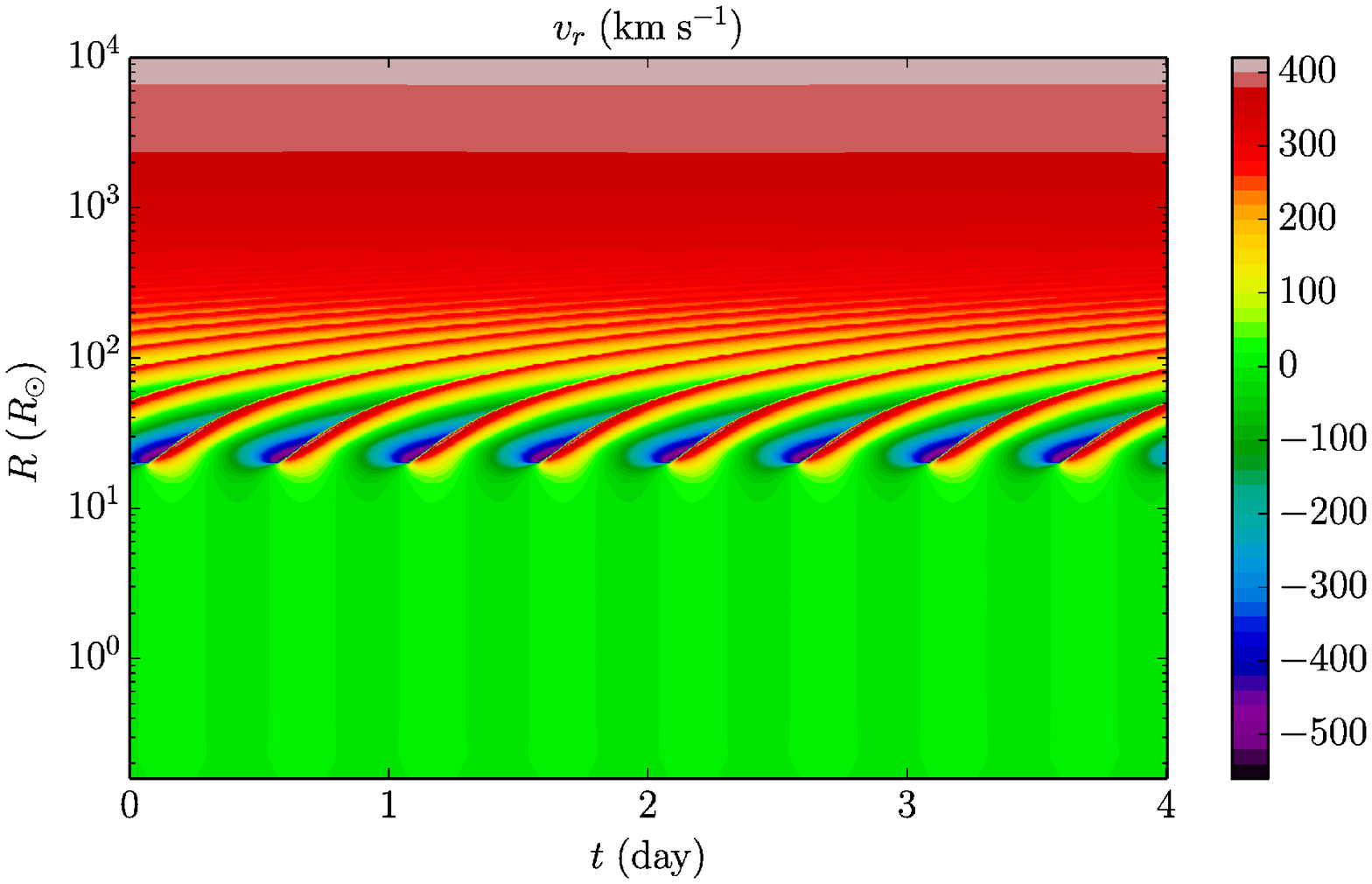}
\caption{
A panoramic view of the pulsations as a function of radius and time, after $t=10^9 \rm{s}$, for run 1 (left panel), and run 4 (right panel).
Positive velocities indicate flow in the positive radial direction.
}
\label{fig:velx_t_r}
\end{figure*}

We analyze the pulsation by restarting run 1 at $t=10^9 ~\rm{s} \simeq 30 \rm{yr}$, and running it with higher sampling.
In order to know the required sampling rate, to avoid any aliasing, we take the maximum Brunt-V\"{a}is\"{a}l\"{a} frequency at $t=10^9 \rm{s}$,
$N_{\rm{BV}} \simeq 3 \times 10^{-4} \rm{Hz}$, which corresponds to $\sim 6$ hours.
The Nyquist sampling rate (maximum allowed sampling rate) to avoid aliasing is therefore $3$ hours.
We restart the run with high sampling rate and let it run for $\sim 4 ~\rm{days}$.
We then slice the star into layers and plot the radial velocity of each layer as a function of time, as shown in Fig. \ref{fig:velx_t} and Fig. \ref{fig:velx_t_r}.
Note that positive velocities indicate flow in the positive radial direction.
We can identify a number of radial regions:
\begin{enumerate}
\item The internal part of the star from the center up to $\sim 12~R_\odot$ is almost not pulsating at all and has a radial velocity close to $v_r=0$.
\item At $\sim 12~R_\odot$ a transition begins into strong radial pulsations, with amplitudes increasing up to a peak at $\sim 20~R_\odot$.
It is also seen that at different radii the pulsations have a phase difference.
The phase difference is gradually increasing with radius.
\item After $\sim 40~R_\odot$ there is a transition region where there are variations in the radial velocity as a function of time,
but not periodic pulsations.
\item At $\sim 400~R_\odot$ the pulsations almost completely stop, and the wind has only a radius dependent velocity profile, with an acceleration trend.
\end{enumerate}
We do not predict observable rapid pulsations, since the amplitude goes to close to zero at the surface of the star,
and since the waves can become non radial and incoherent in three dimensions.

As mentioned above, the region of the star at a few $\times 10~R_\odot$ shows strong decline in $L_{\rm{rad}}$.
As shown in Fig. \ref{fig:velx_t_r}, this is the same regions where stellar pulsation is strongest.
In this region pulsation transforms energy carried by photons to kinetic energy,
generating the wind.

\begin{figure*}
\includegraphics[trim= 0.0cm 1.0cm 1.6cm 2.8cm,clip=true,width=0.5\textwidth]{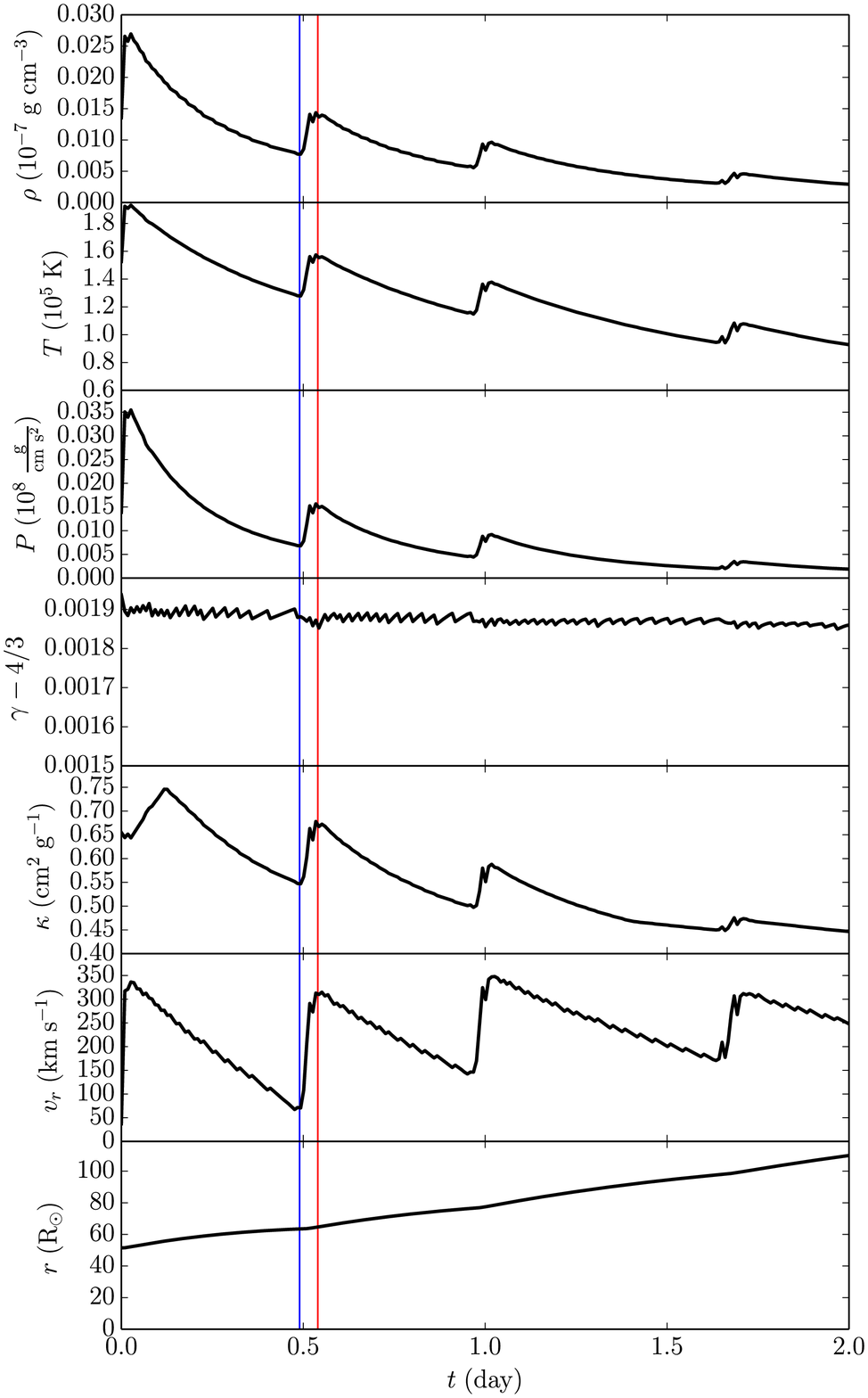}
\includegraphics[trim= 0.0cm 1.0cm 1.6cm 2.8cm,clip=true,width=0.5\textwidth]{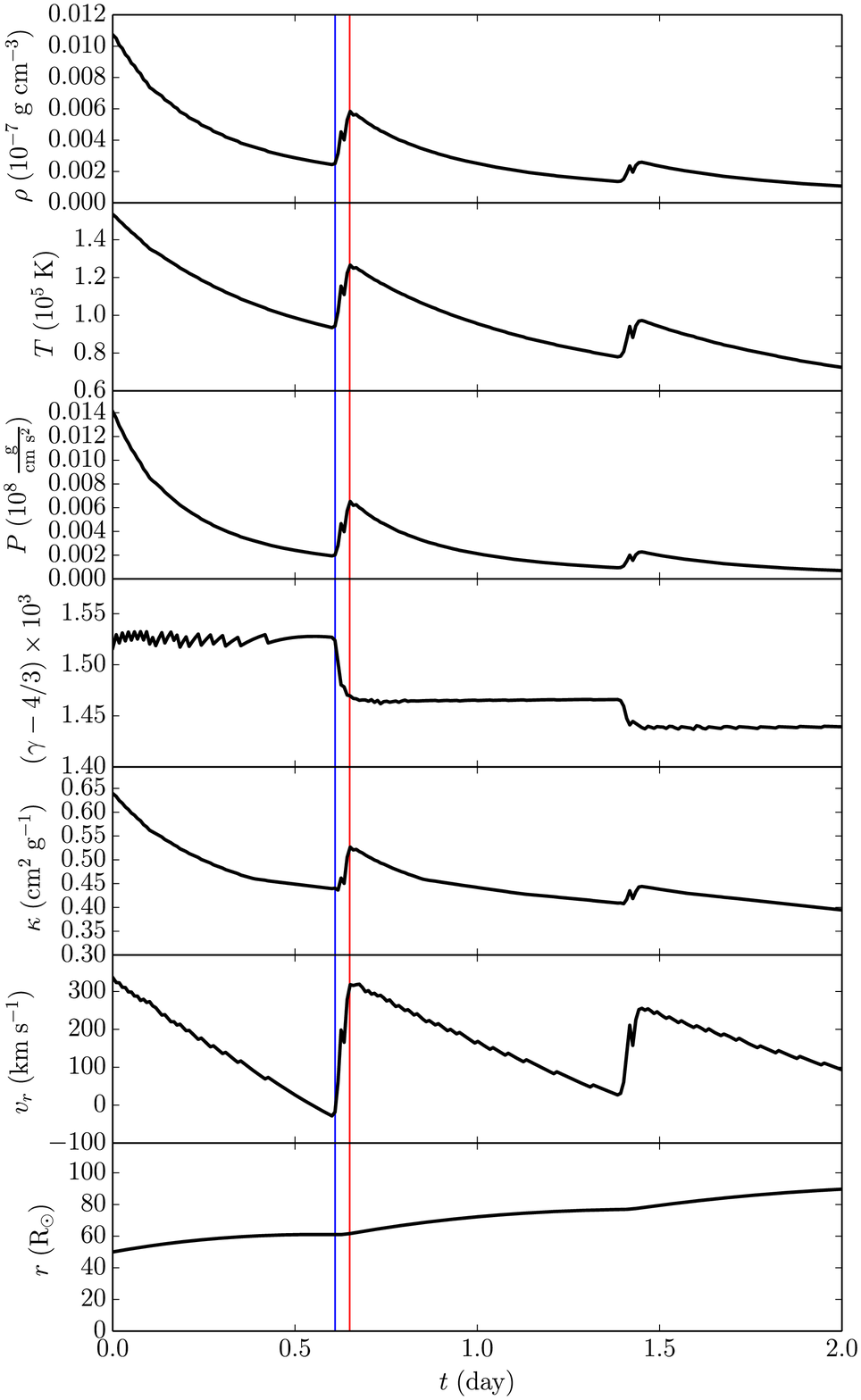}
\caption{
The density, temperature, pressure, adiabatic index, opacity, radial velocity, and radius of a gas element starting at $r=50~R_\odot$
as a function of time, after $t=10^9 \rm{s}$, for run 1 (left panel), and run 4 (right panel).
Note the difference in scale between the two panels.
Positive velocities indicate flow in the positive radial direction.
The vertical lines mark the density peak (red) and minimum (blue).
It is evident that when the pressure increases the opacity increases. This indicates that the $\kappa$-mechanism is driving the pulsations.
}
\label{fig:properties_lagrangian}
\end{figure*}

The \texttt{FLASH} code has an Eulerian grid.
In order to know what happens to a layer of the star as a result of the pulsation we need to follow an element of gas in its trajectory, i.e., in a Lagrangian way.
We therefore start at a certain radius and check the properties at some initial time.
Then, we follow the gas element that started at this radius to different radii in the star according to its velocity.
Fig. \ref{fig:properties_lagrangian} presents density, temperature, pressure, adiabatic index, opacity and radial velocity at starting radius $r=50~R_\odot$, as a function of time, after $t=10^9 \rm{s}$, for runs 1 and 4.
We also calculate the quantity $P^{1-\gamma} T^\gamma$, which for adiabatic pulsations should remain constant with time.
We find that $P^{1-\gamma} T^\gamma$ varies in a periodic way with the same period of the other quantities -- so the pulsations are not adiabatic.
This is not surprising since mass is lost from the system.
An important feature we see in both panels of Fig. \ref{fig:properties_lagrangian}, is that when the pressure increases there is an increase in opacity.
This increase in opacity acts to extract radiation energy from the inner parts of the star, essentially the $\kappa$-mechanism.

\begin{figure*}
\includegraphics[trim= 0.3cm 0.6cm 1.8cm 2.2cm,clip=true,width=0.5\textwidth]{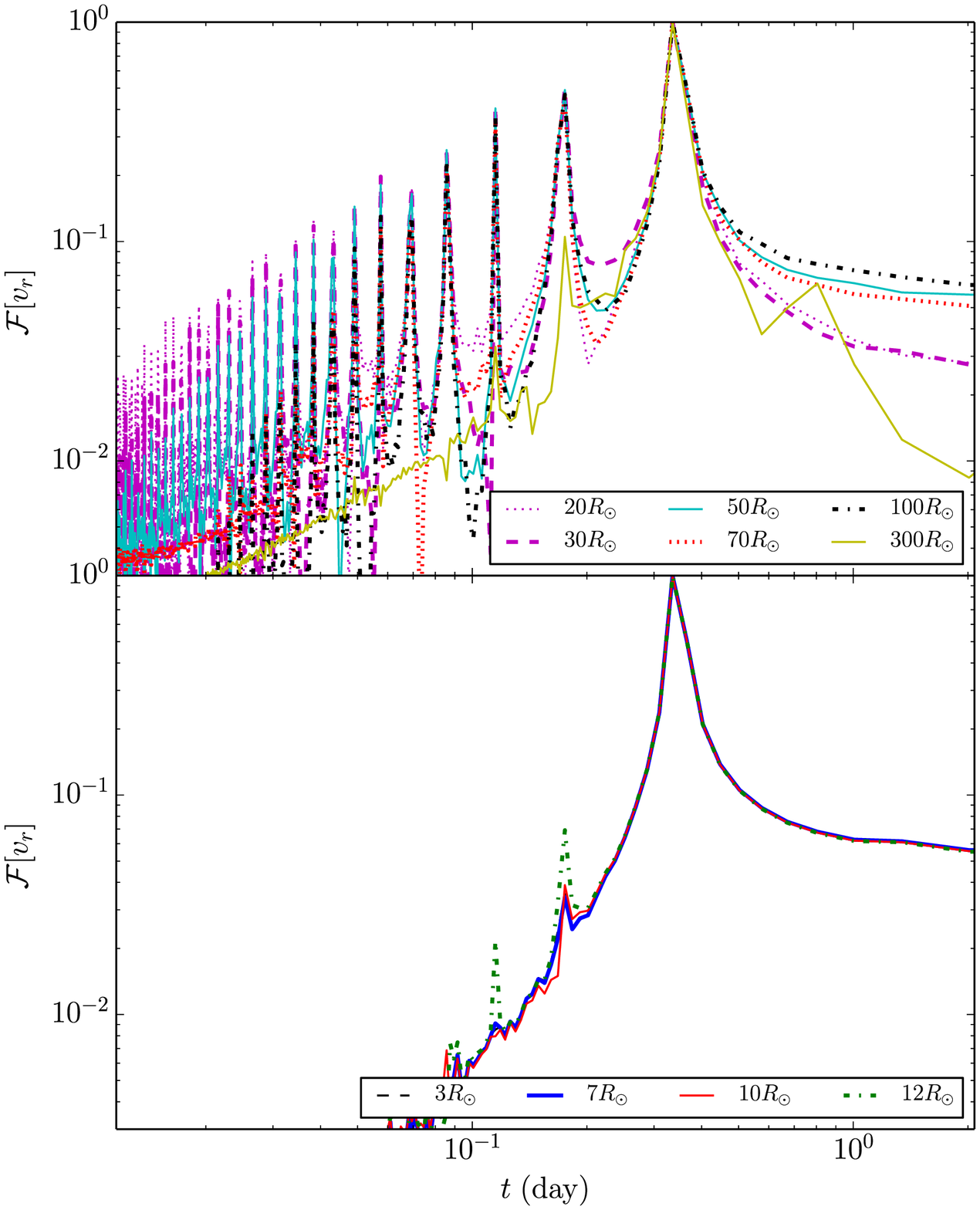}
\includegraphics[trim= 0.3cm 0.6cm 1.8cm 2.2cm,clip=true,width=0.5\textwidth]{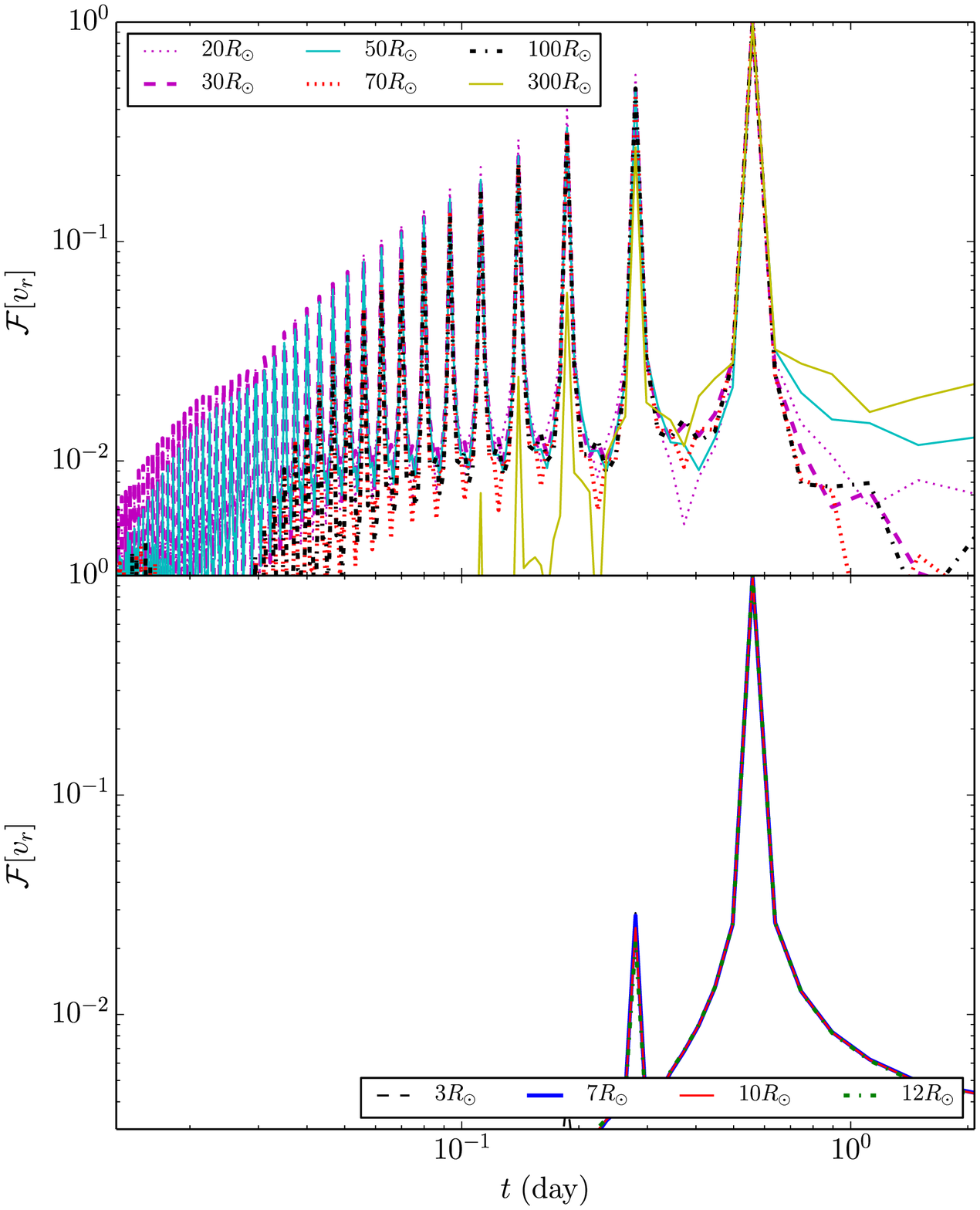}
\caption{
Fourier analysis of the pulsations shown in Fig. \ref{fig:velx_t}.
The ordinate is the normalized Fourier transform of the radial velocities, and the abscissa is the time scale for the pulsation.
Left Panel: Run (1). A pronounced peak is seen at $t_p=0.34$ days.
Right Panel: Run (4). A pronounced peak is seen at $t_p=0.45$ days.
}
\label{fig:pulsations_fourier}
\end{figure*}

Fig. \ref{fig:pulsations_fourier} shows the Fourier transform of the radial velocity as a function of time-scale, at different radii.
We find a pronounced peak at $t_p=0.34~\rm{days}$.
Note that there is a phase difference between the pulsations at different radii,
that is not seen in the Fourier space but clearly seen in the velocity space (Figs. \ref{fig:velx_t} and \ref{fig:velx_t_r}).
Other modes are also seen with smaller time scales, especially for the radii where strong pulsation exists.
At larger radii there is no evident pulsation time scale smaller than $4~\rm{days}$.

Runs 2 and 3 used method 1. The only change was the amount of mass that was removed from the star.
The result are shown in Table \ref{Table:compareruns}.
We see a different behavior between run 2 (where $3 ~M_\odot$ is removed) and runs 1 and 3 (where $6 ~M_\odot$ and $9 ~M_\odot$ are removed, respectively).
Runs 1 and 3 show the same qualitative behavior: a strong mass loss followed by a long duration plateau after which there is a decline accompanied by fluctuations.
Note also that run 3, even more extreme than run 1, did not result in much more total mass ejected.
This is a consequence of the star's initial structure, having a steep gravitational binary energy below the outer $\sim 30 ~M_\odot$.
\begin{figure*}
\includegraphics[width=0.89\textwidth]{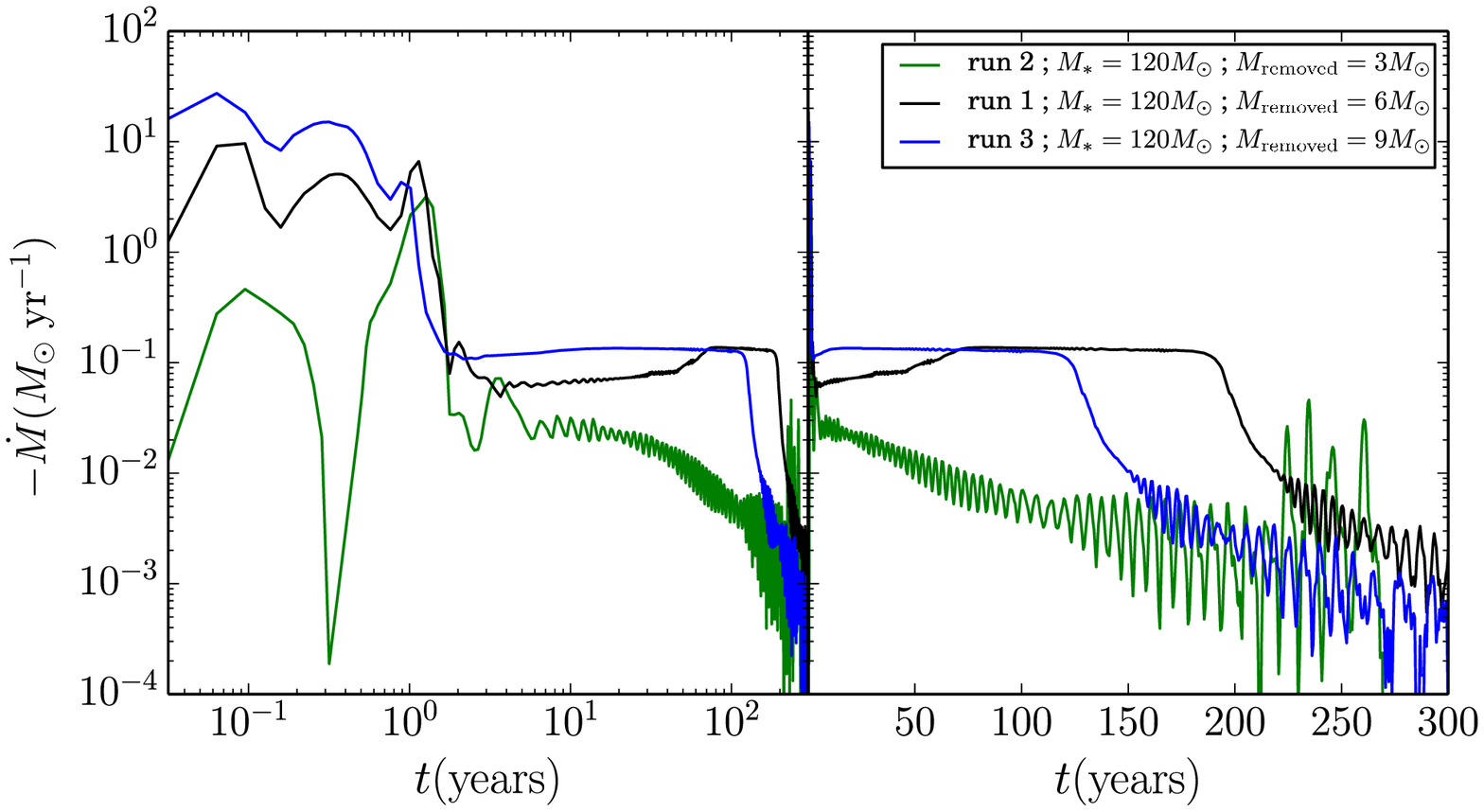}
\caption{
The mass-loss rate as a function of time for runs using the $120 ~M_\odot$ stellar model, and method 1 (runs 1--3).
After a long time in runs 2 and 3 the temperature in the outer parts of the grid went below the $1000 \K$ threshold our code can treat,
and therefore the runs were stopped.
}
\label{fig:time_functions_comparison_120}
\end{figure*}

Run 4 uses the same $120 ~M_\odot$ initial stellar model,
but instead of removing mass, we take the energy required to do so and extract it from the inner layers (method 2 in section \ref{sec:methods:Eruption}).
Specifically, we reduce the thermal energy of each mass element by a constant small fraction,
and deposit that energy in a narrow band of outer layers.
The latter region was adjusted to coincide with the deepest layers that were eventually ejected ($r=10.5 ~R_\odot$).
Its width was arbitrarily chosen to have a width of $3 ~R_\odot$, and the energy was deposited in the those layers weighted by layer's mass.
Narrowing or widening the band of layers (up to a factor of 3) did not significantly alter the results for $t \gtrsim 1$ yr.
The right panel of Fig. \ref{fig:LBV2} shows the properties of the star after extracting energy from the core to a layer, according to method 2, and adding an extension.
The left panels of Fig. \ref{fig:time_functions_multiplot} shows the properties of the star as a function of time.
While the mass and its derivative are calculated unambiguously, the radius, temperature and luminosity are the estimated at $\tau=3$.
These are not actual photosphere values.
The $\tau=3$ location changes rapidly as a function of time.
However, this does not mean that the stellar photosphere is so variable.
Most of the mass is lost in the first $\sim 3 ~\rm{yr}$ of the simulation. This corresponds to the eruption itself, while later times show the recovery of the star from the eruption.

The right/second panels of Figs. \ref{fig:time_functions_multiplot} and \ref{fig:tau} -- \ref{fig:pulsations_fourier} show the results of run 4.
Fig. \ref{fig:multiplot} shows that in run 4 the star is dominated by radiation pressure.
In the envelope, after the eruption only radiation pressure plays a role in pushing the wind; gas pressure is negligible.
As seen in Fig. \ref{fig:Luminosities}, the outer luminosity is radiative luminosity, and the convective luminosity is negligible (this is true for run 1 and all other runs studied here).

Fig. \ref{fig:velx} and Fig. \ref{fig:velx-vesc} show us that the wind velocities that are reached for run 4 are comparable of run 1, reaching $\sim 400 ~\rm{km~s^{-1}}$.
Run 4 has a higher $N_{BV}$ shown in Fig. \ref{eq:N_BV}, but not by much.
The Fourier analysis in Figs. \ref{fig:velx_t} and \ref{fig:velx_t_r} shows that qualitatively the pulsations for both methods are similar,
though somewhat higher frequency for method 1.
The difference in the frequency are not related to the different masses that the two methods have, as at the time analyzed ($t=10^9~\rm{s}$) their masses are similar
$\sim 110 ~M_\odot$.
This is despite run 1 having much lower mass at later times due to the strong runaway mass loss.
In fact the frequencies only differ by a factor of up to $\sim1.5$, and considering the slightly different properties (density, temperature) such a small change is not surprising.
As noted earlier, the waves in a 3D model will surely be less coherent than those in our figures, and hence more chaotic.

\begin{figure*}
\includegraphics[width=0.89\textwidth]{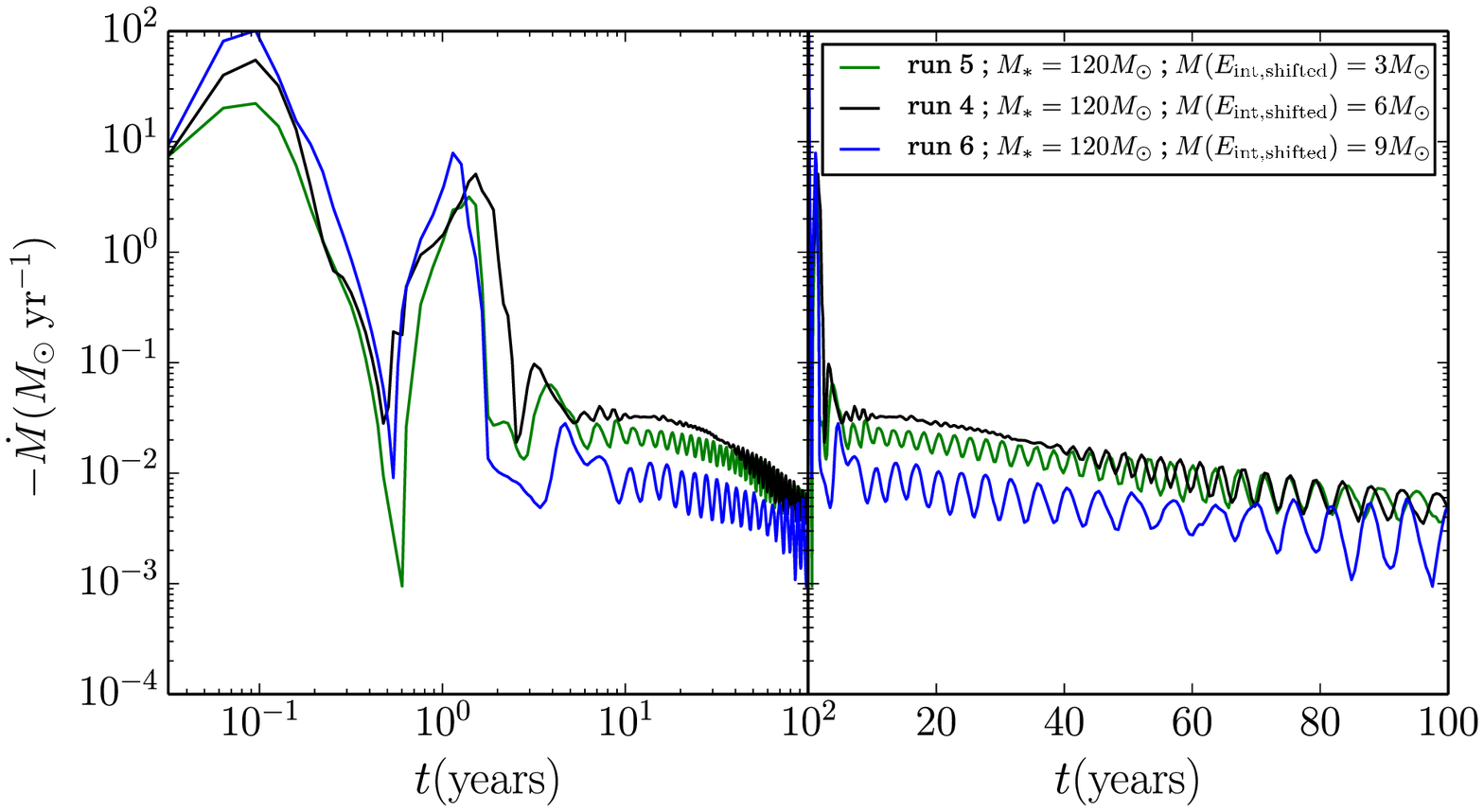}
\caption{
The mass-loss rate as a function of time for runs using the $120 ~M_\odot$ stellar model, and method 2 (runs 4--6).
}
\label{fig:time_functions_comparison_120_ES}
\end{figure*}

For method 2 we ran three runs for the $120 ~M_\odot$ stellar model with different energy extracted from the core, which is equivalent to the energy required for ejecting a certain amount of mass, as indicated in Table \ref{Table:compareruns} (referred to as ``equivalent energy shifted'').
The total mass lost is larger then this amount, as other sources of energy that operate after the eruption remove the extra mass.
The properties and the evolution with time of the three runs are very similar qualitatively.
Fig. \ref{fig:time_functions_comparison_120_ES} shows a comparison of the mass-loss rate as a function of time of the three runs.
Most of the mass is lost in the first $\sim 3 ~\rm{yr}$ of the simulation.
We can clearly see a trend in the mass-loss rate during that phase, with the more energy extracted from the core the higher the mass-loss rate during that time.
At later time there is no clear trend in our results.

\begin{figure*}
\includegraphics[width=0.89\textwidth]{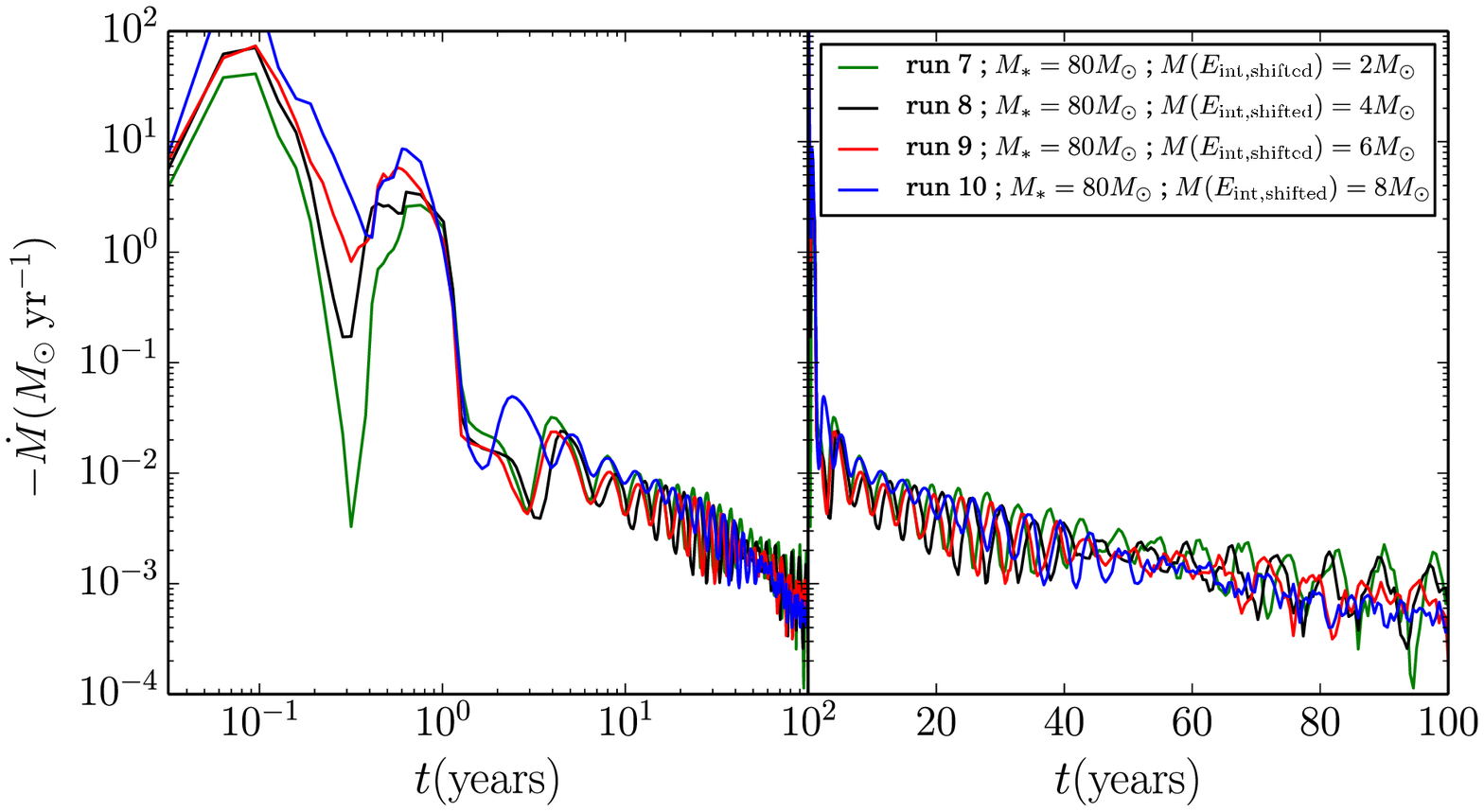}
\caption{
The mass-loss rate as a function of time for runs using the $80 ~M_\odot$ stellar model, and method 2 (runs 7--10).
}
\label{fig:time_functions_comparison_80_ES}
\end{figure*}

We ran four more runs for a different stellar model with $M=80 ~M_\odot$, following the same procedure used to obtain the $120 ~M_\odot$ stellar model.
The four runs differ by the energy extracted from the core, which is equivalent to the energy required for ejecting a certain amount of mass, as indicated in Table \ref{Table:compareruns}.
Fig. \ref{fig:time_functions_comparison_80_ES} shows a comparison of the mass-loss rate as a function of time of the four runs.
The same trends seen for the $120 ~M_\odot$ stellar model are also observed here.

\section{Summary and Discussion}
\label{sec:Summary}

In each model described here, a giant VMS eruption tends to prolong
itself for reasons that do not depend much on the instability
that started it. The later aftermath includes a very strong wind
($\dot{M} \gtrsim 10^{-3} ~M_\odot~\rm{yr^{-1}}$)
which decays in one or two centuries. This outcome appears
physically reasonable, since the initial mass-loss obviously
produces a state of serious internal disequilibrium.
(For example, a large amount of radiation energy suddenly
finds itself less trapped than it was before.) Moreover, the
picture also appears consistent with the only well-observed
giant-eruption survivor, $\eta$ Car. The main cause for doubt
is that a 2D or 3D outburst might, in principle, abate sooner;
but we have not noticed any simple or obvious reasons why
this should be the case, apart from factors of order 2.
Recently \citet{Quataert2015} have, like us,
employed 1D calculations for a related problem.
Our models were planned to serve as a 1D reconnaissance of
the problem, pending 2D and 3D work which will be
far more difficult. In that sense, perhaps the most important
result here is that 3D models now seem more promising
in view of the quantitative 1D behavior.

When tracking the propagation of gas elements, there is a correlation between the compression of the gas and the increase in opacity.
We therefore classify the resulting pulsations as from $\kappa$-mechanism.
The so called ``strange-mode'' instability was suggested as a possible driving force for winds and even giant eruptions (e.g. \citealt{Sterken1996}; \citealt{vanGenderen1995}; \citealt{cox1997}; \citealt{Guzik1997}, \citeyear{Guzik2012}).
Exploring the existence or absence of strange modes in the pulsations we obtained is beyond the scope of the present paper.

The amplitude of the pulsations is almost certainly exaggerated, simply because these are 1D calculations.
In 3D the same instability exists but would not be radially coherent.
High amplitudes may exist locally, but their directions and phases would vary with location in the envelope.
Therefore the averaged mass loss rate is credible, but there is no reason to expect it have such obvious, coherent global pulsations.

To check the effects of rotation, we also ran a \texttt{MESA} model where the initial rotation rate is $\Omega = 40\% $ of the critical maximum.
Adding rotation had little effect when studying the eruption, as the outer shells, which possess most of the angular momentum, were removed.

Our simulations do \emph{not} explain the major fluctuations that $\eta$ Car has exhibited since its 1830--1860 Great Eruption.
A second outburst was observed around 1890,
a sort of change-of-state rapidly occurred around 1940--1950, and
another rapid transition began in the 1990's (\citealt{Humphreys1999}; \citealt{Martin2006}; \citealt{Mehner2010}; \citealt{HumphreysMartin2012} and
other references therein) -- almost hinting at a 50-year quasi-periodicity.
Many authors have speculated that these peculiarities have something
to do with the companion star, because tidal effects may be
appreciable in the primary's wind-acceleration zone. But such
effects are much weaker in the star's \emph{interior},
$r < 150 ~R_\odot$.
It will be interesting to learn whether 3D simulations behave unsteadily.

\begin{figure*}
\includegraphics[trim= 0.4cm 0.1cm 0.9cm 0.8cm,clip=true,width=0.5\textwidth]{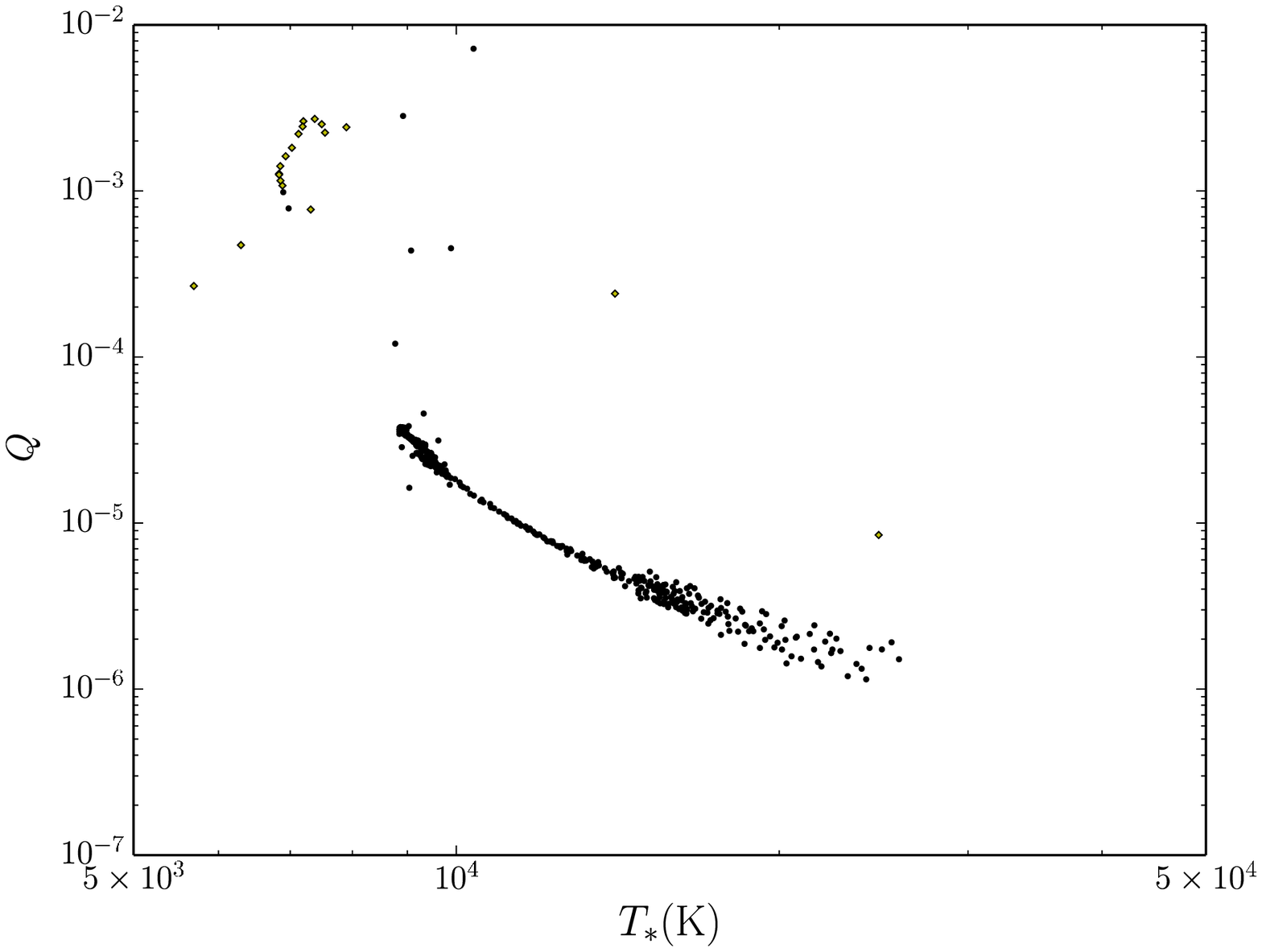}
\includegraphics[trim= 0.4cm 0.1cm 0.9cm 0.8cm,clip=true,width=0.5\textwidth]{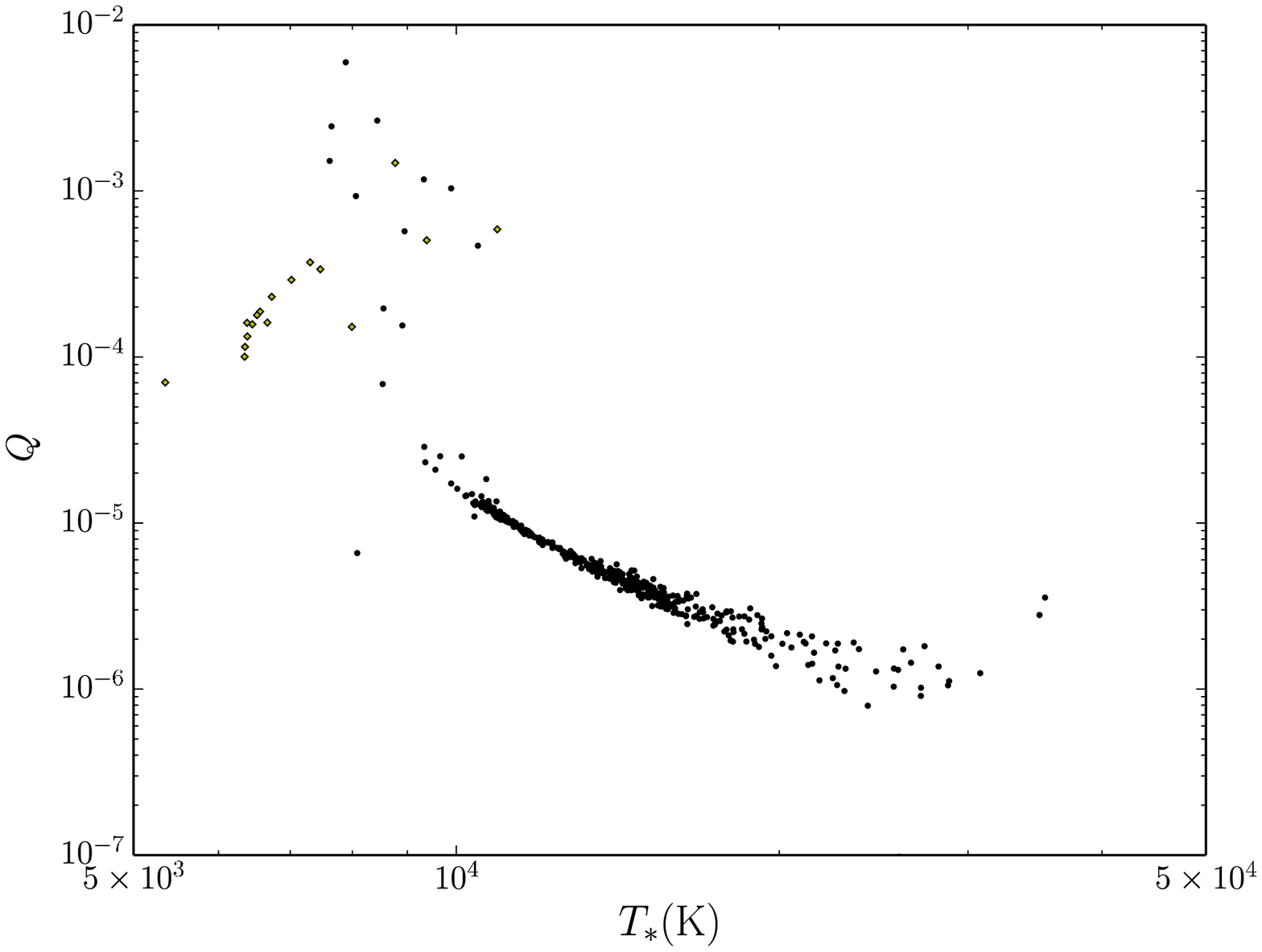}
\caption{
The quantity $Q(T)$ (equation \ref{eq:Q}) for different times during the simulation for run 1 (left) and run 4 (right).
The yellow diamonds are early times in the simulation and the values of T are not accurate.
The black dots are from times later in the simulation.
We get that the simulation gives $T > 7500 \K$, the same results calculated analytically by \cite{Davidson1987}.
}
\label{fig:Q_T_1}
\end{figure*}

Our simulations also provide information about how the giant eruption may have looked.
\cite{Davidson1987} calculated the radiation temperature of luminous stars undergoing eruptions.
He found that the temperatures do not fall much below $\sim 6000 \K$ even for enormous mass-loss rates.
\cite{Davidson1987} defined the quantity $Q$ as
\begin{equation}
Q(T)=\left(\frac{\dot{M}}{M_\odot~\rm{yr^{-1}}}\right)   \left(\frac{v}{\rm{km~s^{-1}}}  \vphantom{\frac{\dot{M}}{M_\odot}}  \right)^{-1}  \left(\frac{L}{10^6 L_\odot}  \vphantom{\frac{\dot{M}}{M_\odot}}  \right)^{-0.7},
\label{eq:Q}
\end{equation}
and showed that there is an asymptote close to $\sim 6000 \K$.
This is analogous to the Hayashi Limit since it results from the rapid decrease of opacity below $7000 \K$.
For run (4) we calculate the value of $Q(T)$ for different times during the simulation.
Fig. \ref{fig:Q_T_1} shows the results.
The yellow diamonds are early times in the simulation and the values of $T$ are not accurate.
The black dots are from times later in the simulation.
We find that the simulation gives $T > 6000 \K$, the same results calculated analytically by Davidson
\footnote{
\cite{Rest2012} suggested that 6000 K is too high
based on light-echo spectra, but they relied on
comparisons with stellar atmospheres rather than opaque
outflows.  A wind with a $6000 \K$ photosphere can produce
relatively cool spectral features, cf.\ \cite{DavidsonHumphreys2012b}. }.
After the eruption, we found the stellar temperature is above $10^4 \K$, so it stays on the blue side of the HD limit on the HR diagram.

\begin{figure}
\includegraphics[trim= 0.4cm 0.3cm 0.6cm 1.8cm,clip=true,width=0.48\textwidth]{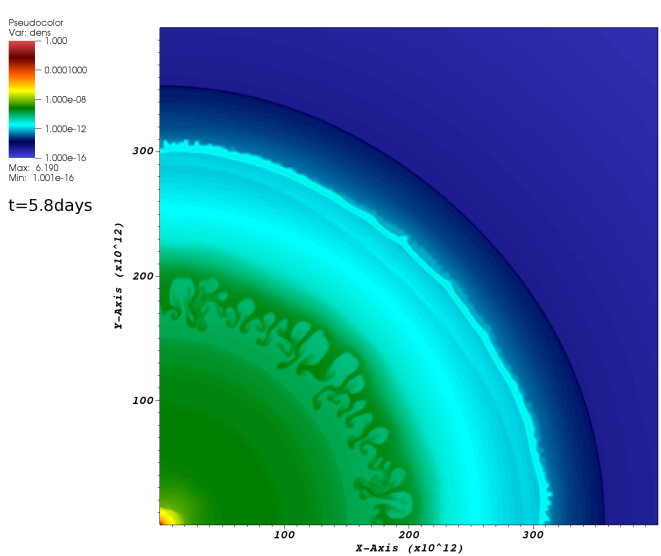} 
\caption{
Preliminary 2D simulation results. The figure shows method 2 density map taken $5.8$ days after the eruption.
A layer experiencing the Richtmyer-Meshkov instability is seen at $r \simeq 1.6$--$2.1 \times 10^{14} \rm{cm}$.
This causes strong mixing of elements from the core to outer layers.
It therefore explains why the ejecta of $\eta$ Car is Nitrogen rich. Nuclear
processed material from the core reaches the outer layers and eventually is ejected from the star.
}
\label{fig:RMI}
\end{figure}

Though 2D and 3D simulations are computationally very expensive, and will be investigated in a future paper,
we have studied the same problem in 2D for a short time interval ($\sim 24$ years).
The case studied is the same as in run 4.
The results we obtained are qualitatively the same.
We do find an interesting result that could not be obtained in 1D simulations.
Under the assumption that the source of the eruption is the
core (method 2), the eruption causes a Richtmyer-Meshkov instability in the star
in the first few days after the eruption (Fig. \ref{fig:RMI}).
This causes strong mixing of elements from the core to outer layers.
It may therefore explain why the ejecta of $\eta$ Car is Nitrogen rich.
Nuclear processed material from the core reaches the outer layers and eventually is ejected from the star.

\begin{figure}
\includegraphics[trim= 0.4cm 0.0cm 0.6cm 0.1cm,clip=true,width=0.47\textwidth]{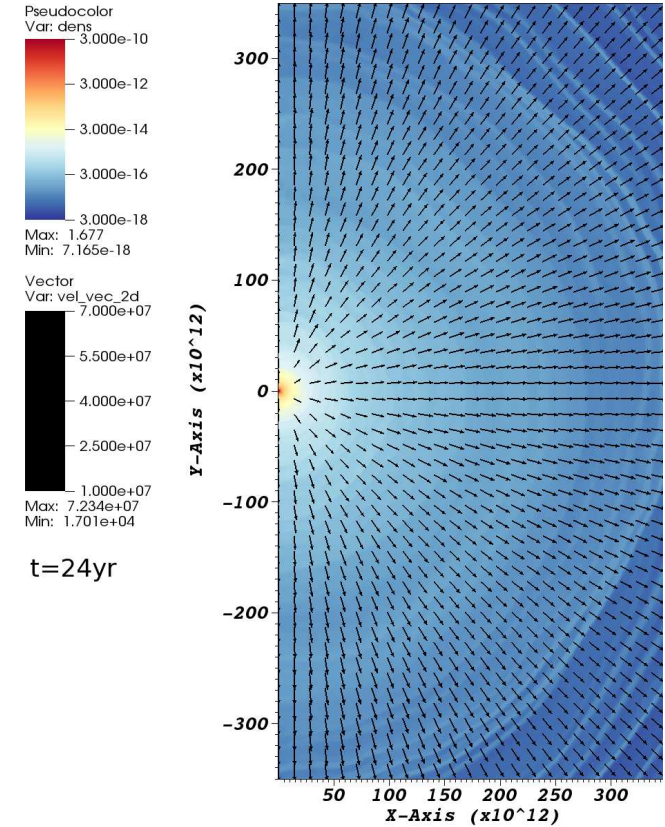}
\caption{
Preliminary 2D simulation results. The figure shows method 2 density map taken $24$ years after the eruption.
The density scale was truncated to show features in the envelope (the central density is $\sim 2 ~\rm{g~cm^{-3}}$).
The simulation shows radial outflow with ripples as a result of pulsations.
The spatial wavelength of the leading mode is longer than in the 1D simulations.
}
\label{fig:302_1520}
\end{figure}

We also find in our 2D simulation that the star experiences pulsations and ejects mass.
Fig. \ref{fig:302_1520} shows a density map with velocity vectors taken $24$ years after the eruption.
We see similar pulsations as in the 1D simulation, but as expected in the discussion above they are not as intense as in the 1D case.

Full multi-dimensional simulations will be the subject of a future paper.
The tools that we are developing for studying giant eruptions will have other uses in studying other objects that experience large mass loss.

\vspace*{0.5cm}
We acknowledge support provided by National Science Foundation through grant AST-1109394.
This work was carried out in part using computing resources at the University of Minnesota Supercomputing Institute (MSI).
This work used the Extreme Science and Engineering Discovery Environment (XSEDE),
which is supported by National Science Foundation grant number ACI-1053575.
We thank Alexander Heger for enlightening discussions about the simulations.
We thank Joyce Guzik and Ken Chen for helpful suggestions.
We thank an anonymous referee for helpful comments.

\vspace*{1cm}


{}

\end{document}